\documentclass[]{aa}

\usepackage{xcolor}
\usepackage{float}
\usepackage{stfloats}
\usepackage{graphicx}
\usepackage{graphbox}
\usepackage{placeins}
\usepackage{chngcntr}
\usepackage{amsmath}
\usepackage{amssymb}
\usepackage{mathtools}
\usepackage{multirow}
\usepackage{verbatim}
\usepackage[toc]{appendix}
\usepackage{soul}
\usepackage{afterpage}
\usepackage[varg]{txfonts}
\usepackage{JournalAbb}
\usepackage{natbib}
\usepackage[colorlinks,citecolor=blue,urlcolor=blue,bookmarks=false,linkcolor=blue,hypertexnames=true]{hyperref} 
\bibpunct{(}{)}{;}{a}{}{,}
\makeatletter
\renewcommand*\aa@pageof{, page \thepage{} of \pageref*{LastPage}}
\makeatother

\newcommand{\TheTH}{\textsc{The300}}
\newcommand{\XMM}{XMM-\textit{Newton}}
\newcommand{\Chandra}{\textit{Chandra}}
\newcommand{\Planck}{\textit{Planck}}

\defcitealias{Bourdin2017}{B17}
\defcitealias{Kozmanyan2019}{K19}
\defcitealias{DeLuca2021}{DL21}

\begin{document} 

    \title{{\sc\bf The Three Hundred} Project: Validating $H_0$ inference from mock X-ray and millimetre analyses of galaxy clusters}

    \titlerunning{\TheTH{} Project: Validating $H_0$ inference from mock X-ray and millimetre analyses of galaxy clusters}
    
    \author{
        F. De Luca\thanks{\email{federico.deluca@roma2.infn.it}}\inst{\ref{UniR2},\ref{INFN-R2}}
        \and
        H. Bourdin\inst{\ref{UniR2},\ref{INFN-R2}}
        \and
        P. Mazzotta\inst{\ref{UniR2},\ref{INFN-R2}}
        \and
        E. Rasia\inst{\ref{INAF-oat}, \ref{IFPU}, \ref{Umich}}
        \and
        A. Kozmanyan\inst{\ref{UniR2}}
        \and
        W. Cui\inst{\ref{UAM}}
        \and
        M. De Petris\inst{\ref{UniR1}}
        \and
        D. de Andres\inst{\ref{URJC}}
        \and
        G. Yepes\inst{\ref{UAM}, \ref{CIAFF}}
    }
    
    \institute{
        Dipartimento di Fisica, Università di Roma ‘Tor Vergata’, Via della Ricerca Scientifica 1, I-00133 Roma, Italy.\label{UniR2}
        \and
        INFN, Sezione di Roma ‘Tor Vergata’, Via della Ricerca Scientifica, 1, 00133, Roma, Italy.\label{INFN-R2}
        \and
        INAF - Osservatorio Astronomico di Trieste, via Tiepolo 11, I-34131, Trieste, Italy\label{INAF-oat}
        \and
        IFPU, Institute for Fundamental Physics of the Universe, Via Beirut 2, 34014 Trieste, Italy\label{IFPU}
        \and
        Department of Physics; University of Michigan, Ann Arbor, MI 48109, USA\label{Umich}
        \and
        Departamento de Física Teórica, Facultad de Ciencias, Universidad Autónoma de Madrid, 28049 Cantoblanco, Madrid, Spain\label{UAM}
        \and
        Dipartimento di Fisica, Sapienza Università di Roma, Piazzale Aldo Moro 5, I-00185 Rome, Italy\label{UniR1}
        \and
        Nonlinear Dynamics, Chaos and Complex Systems Group, Departamento de Geología, Física y Química Inorgánica, Universidad Rey Juan Carlos, Tulipán s/n, 28933 Móstoles, Madrid, Spain\label{URJC}
        \and
        Centro de Investigación Avanzada en Física Fundamental (CIAFF), Facultad de Ciencias, Universidad Autónoma de Madrid, 28049 Madrid, Spain\label{CIAFF}
    }
    
    \date{Received Month Day, Year; accepted Month Day, Year}

    \abstract{Measurements of thermodynamical quantities in galaxy clusters are differently affected by simplified modelling of radially averaged observables in the X-ray and millimetre bands. This includes assumptions about the cosmological model and the morphology of the cluster intracluster medium (ICM). Within a large sample of clusters extracted from {\sc The Three Hundred} hydrodynamical simulations, we assess the systematic differences expected from the morphological assumptions between ICM temperatures as inferred from X-ray spectroscopy or joint X-ray and millimetre imaging. We find that these differences show a well-defined statistical behaviour that correlates with the cluster dynamical and morphological indicators. We then investigate how joint inferences of cluster temperature profiles, a priori informed by this statistical behaviour, allow us to constrain cosmological parameters inferred from the apparent cluster sizes. Assuming a flat $\Lambda$ cold dark matter ($\Lambda$CDM) cosmology and priors on $\Omega_\mathrm{m}$ and the helium abundance, this method provides us with unbiased estimates of the Hubble constant, $H_0$, characterised with a precision of about $4\%$ and $1.5\%$ for samples of 100 and 1000 clusters, respectively, and ultimately limited by systematic uncertainties of about $0.6$--$0.8\, {\rm km\, s^{-1} Mpc^{-1}}$. This work highlights the potential of joint X-ray and millimetre observations of galaxy cluster samples to place tight constraints on $H_0$.}
    
    \keywords{
        Methods: numerical -- Galaxies: clusters: general --
        Galaxies: clusters: intracluster medium --
        Cosmology: cosmological parameters
    }

    \maketitle

\section{Introduction} \label{sec:intro}

    In modern cosmology, numerical simulations provide an ideal framework for making predictions about current or new cosmological models in controlled environments or for the interpretation of observations \citep[for reviews, see][]{Vogelsberger2020, Angulo2022}. The formation and evolution of galaxy clusters, for example, are influenced by the large-scale structure of the Universe and the physical evolution of their components \citep{Kravtsov2012, Planelles2015}: the dark matter halo and the gravitationally trapped baryons within galaxies and in the hot intracluster medium (ICM). Galaxy clusters can therefore be used as cosmological probes, provided that their structures are accurately modelled \citep[][]{Voit2005, Allen2011, Weinberg2013, Miyatake2025}. This task can be addressed with numerical methods. Indeed, N-body and hydrodynamical simulations have successfully predicted several cluster properties, such as the overall self-similarity of ICM thermodynamic profiles \citep[driven by the hierarchical gravitational growth of the dark matter halo; e.g.][]{Kaiser1986, Vikhlinin2006a, Kravtsov2006, Nagai2007, Arnaud2010, Bohringer2012, Lau2015}. In particular, hydrodynamical simulations have brought to light some limitations of the simplistic $N$-body predictions that change once baryon physics is considered \citep[][]{Battaglia2012, McDonald2017, Truong2018, Ghirardini2019, Gaspari2019}.

    Regarding cluster-based cosmological probes, cosmological models can be constrained by studying the distribution and evolution of galaxy clusters in the Universe. For example, cluster number counts, clustering, and gas fractions can be used to constrain model parameters such as baryon and matter densities, $\Omega_\mathrm{b}$ and $\Omega_\mathrm{m}$, normalisation of the matter power spectrum, $\sigma_8$, as well as the properties of dark matter and dark energy \citep[e.g.][]{Markevitch2004, Vikhlinin2009, Planck2016XXIV, Bocquet2024, Sunayama2024, Ghirardini2024}. However, systematic uncertainties in cluster modelling limit the constraining power of cluster cosmology. For mass-based methods (e.g. number counts and baryon fraction measurements), one major limitation arises from the mass estimates themselves. Cluster masses are generally indirectly inferred from other observables, such as X-ray temperature and luminosity, the Sunyaev-Zel'dovich (SZ) signal, or lensing measurements. These methods may be subject to systematic uncertainties \citep[see][for a review]{Pratt2019}, given by assumptions on the ICM physics and geometry, such as hydrostatic equilibrium, spherical symmetry, and the absence of clumps \citep{Sulkanen1999, Kawahara2008, roncarelli2013, Eckert2015, Lau2020}. To mitigate these effects, analyses are often restricted to subsamples of the so-called ‘relaxed’ systems, which are expected to be closer to hydrostatic equilibrium and exhibit nearly spherical morphologies. However, genuinely relaxed systems are only a minority of current samples, and residual systematics may still affect the analyses. Hydrodynamical simulations have been widely used to study and model these systematics. Inhomogeneities, triaxiality, active galactic nuclei (AGN) feedback, and non-thermal pressure support have all been shown to induce mass biases, which may also depend on the redshift, mass, or dynamical state \citep{Gaspari2014, Henson2017, Ansarifard2020, Grandis2021, Gianfagna2022, Rasia2025}.
    
    Other cluster-based probes are based on standard ruler methods \citep{Cowie1978, Silk1978, Cavaliere1979}. These techniques infer the angular diameter distance by combining millimetre and X-ray observations of galaxy clusters, from which one can obtain the Hubble parameter, $H_0$. Both signals originate from the optically thin ICM, providing line-of-sight (LOS) integrated information on the cluster plasma. At X-ray wavelengths, the observed emission is due to thermal bremsstrahlung and metal lines \citep[see][for a review]{Bohringer2010}, while at millimetre wavelengths the signal is produced by the inverse Compton scattering of the cosmic microwave background photons with the free electrons in the ICM, known as the thermal SZ effect \citep{Sunyaev1972, Sunyaev1980}
    \begin{align}
        &\Sigma_\mathrm{X} = \frac{1}{4\pi(1+z)^4}\int n_\mathrm{p} n_\mathrm{e} \Lambda_\mathrm{X}(T_\mathrm{e}, Z) \, dl,
        \label{eq:X_sb} \\
        &I_\mathrm{SZ} = \frac{\Delta T_{\rm CMB}}{T_{\rm CMB}} = y \cdot g(\nu),
        \label{eq:I_y}  \\
        &y = \dfrac{\sigma_T}{m_\mathrm{e} c^2}\int n_\mathrm{e} kT_\mathrm{e} dl,
        \label{eq:y}
    \end{align}
    where $z$ is the cluster redshift, $\sigma_T$ is the Thomson cross section, $k$ is the Boltzmann constant, $m_\mathrm{e}$ is the electron rest mass, and $\nu$ is the observing frequency. The function $\Lambda_\mathrm{X}(T_\mathrm{e}, Z)$ is the X-ray cooling function, where $Z$ denotes the ICM metallicity, $g(\nu)$ describes the spectral shape of the SZ signal, and $y$ is the Compton parameter. Throughout the LOS, the effect of SZ depends linearly on the electron density, $n_\mathrm{e}$, the temperature $T_\mathrm{e}$, and hence on the pressure. The X-ray emission depends on the product of the proton ($n_\mathrm{p}$) and electron densities, and scales approximately as $n_\mathrm{e}^2$ for a given ICM chemical composition. The LOS integrals in Eqs.~\eqref{eq:X_sb} and~\eqref{eq:y} depend on the angular diameter distance of the cluster: $dl = D_\mathrm{A} d\zeta$, with $\zeta$ denoting the LOS angular size \citep[][]{Birkinshaw1999, Carlstrom2002}. As a result, the quantity $y^2/\Sigma_\mathrm{X}$ provides a density-weighted measure of the LOS size of the cluster. With an accurate model of the ICM structure, we can then relate the apparent size of the cluster to $y^2/\Sigma_\mathrm{X}$, and thus measure $D_\mathrm{A}$ \citep[e.g.][]{Reese2004, Schmidt2004, Bonamente2006}.

    As detailed in \citet[][hereafter \citetalias{Kozmanyan2019}]{Kozmanyan2019}, standard ruler methods can constrain $H_0$ not only from observed quantities ($y^2/\Sigma_\mathrm{X}$), but also from the ICM thermodynamic that can be extracted from X-ray and millimetre observations, such as the ICM pressure or temperature. In the ideal case, when the cosmology is well known as the cluster dynamical state, we expect no discrepancies between the SZ and X-ray derived profiles. However, in real cases, discrepancies arise as a result of assumptions made about the cluster structure and the underlying cosmology. If these two sources of bias are not correlated, the discrepancy factor can be written as $\eta = \mathcal{C} \cdot \mathcal{B}$ \citepalias[for more details, see][]{Kozmanyan2019}, where $\mathcal{B}$ and $\mathcal{C}$ are the overall bias terms given by the cluster and cosmological  modelling, respectively. 
    
    Therefore, by comparing the X-ray and SZ thermodynamic properties, we can improve our models of the cluster structure and of the cosmological framework \citep[e.g.][]{Sereno2018, Ettori2020, Kim2024, Chappuis2025}. In \citetalias{Kozmanyan2019}, hydrodynamical simulations were used to model the distribution of $\mathcal{B}$ and include this information as informative priors in their Bayesian analysis, applied to data from \XMM{} and \Planck{} satellites. Following this approach, the analysis is not limited only to small relaxed samples, thereby increasing the effective sample size. In addition, by relying on thermodynamic profiles rather than integrated quantities, the method exploits the radial information contained in the X-ray and SZ signals as a cosmological probe. A similar strategy to that of \citetalias{Kozmanyan2019} was adopted by \citet{Wan2021} for a relaxed subsample (for which $\mathcal{B} \sim 1$) observed with \Chandra{}, \Planck{}, and Bolocam. Since current X-ray telescopes show systematic differences in cluster temperature measurements \citep[e.g.][]{Reese2010, Nevalainen2010, Schellenberger2015, Migkas2024}, \citet{Wan2021} also included an empirical prior on the X-ray temperature calibration bias, obtained by combining weak-lensing and X-ray mass estimates with simulation-based expectations for the hydrostatic mass bias.

    In this work, we build on and expand the simulation analysis of \citetalias{Kozmanyan2019} to a larger sample of synthetic clusters from {\sc The Three Hundred} project (hereafter \TheTH{}). In particular, we derive new measurements of the $\mathcal{B}$ distribution and validate a new cosmological Bayesian pipeline that also includes the morphological state of the clusters in the modelling of the informative prior on $\mathcal{B}$. The paper is organised as follows. In Section~\ref{sec:sim}, we present the simulation suite and the cluster catalogue. In Section~\ref{sec:meth}, we describe our methodology, with the characterisation of the thermodynamic profiles, the morphological state, and the modelling of the informative priors. The $\mathcal{B}$ distribution and the validation of the informative priors in the cosmological pipeline are then presented in Section~\ref{sec:res} and discussed in Section~\ref{sec:disc}. Finally, our conclusions are summarised in Section~\ref{sec:conc}. Throughout the paper, ‘$\log$’ and ‘$\ln$’ denote the base $10$ and natural logarithms, respectively.

\section{The simulation} \label{sec:sim}

    \TheTH{} project\footnote{\url{https://www.nottingham.ac.uk/astronomy/The300/}} consists of zoom-in hydrodynamical simulations \citep{Cui2018} based on cosmology consistent with the results of \citet{Planck2016XIII}, with $h=0.678$ \citep[where $h$ is defined as $H_0 = 100 \,h \, {\rm km \, s^{-1} Mpc^{-1}}$, e.g.][]{Croton2013}, $\Omega_\mathrm{m}=0.307$, and $\Omega_\Lambda=0.693$. In particular, regions of $15h^{-1} {\rm Mpc}$ in radius around the $324$ most massive clusters identified at $z=0$ in the MultiDark-Planck2\footnote{\url{https://www.cosmosim.org/metadata/mdpl2/}} \citep{Klypin2016} dark matter-only simulation were re-simulated adding baryons (in the highest resolution volume, the gas and dark matter particles have masses equal to $m_{\rm gas} = 2.36 \times 10^8 h^{-1} {\rm M_\odot}$, $m_{\rm DM} = 1.27 \times 10^9 h^{-1} {\rm M_\odot}$) and with different hydrodynamical codes. 
    
    Currently, \TheTH{} collects the results produced with {\sc gadget-music} \citep{Sembolini2012}, {\sc gadget-x} \citep{Rasia2015}, and {\sc gizmo-simba} \citep{Cui2022} suites. For this work, we considered the {\sc gadget-x} run, since it has been shown to reproduce well several observed cluster properties, such as emission measure profiles \citep{Bartalucci2023} and morphological indicators \citep{Campitiello2022}, the shape and scatter of temperature profiles \citep{Rossetti2024}, and the level of temperature inhomogeneities \citep{Lovisari2024}.

\subsection{The cluster sample} \label{ssec:sample}

    \begin{figure*}
        \includegraphics[width=2\columnwidth]{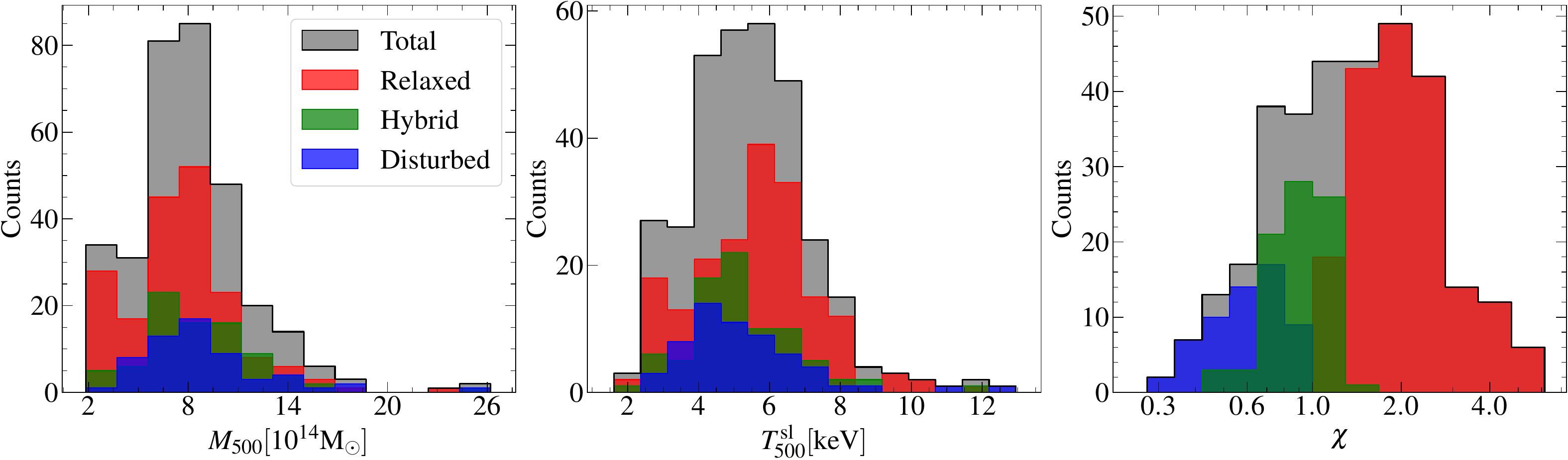}
        \caption{Overall distributions (grey histograms) of $M_{500}$ (left panel), the spectroscopic-like temperatures (centre), and the dynamical indicator $\chi$ (right). Relaxed, hybrid, and disturbed subsamples are shown with red, green, and blue colours, respectively.}
        \label{fig:sample}
    \end{figure*}

    Our sample is carried out with {\sc gadget3} \citep[an updated version of {\sc gadget2};][]{Springel2005}, which adopts a tree-PM gravity solver and an improved smooth-particle-hydrodynamics (SPH) code \citep{Beck2015}. The simulation includes gas cooling, star formation, and feedback from AGN and stars, implemented similarly to \citet{Rasia2015}. For this study, we selected $324$ haloes identified at redshift zero (snapshot 128) with the Amiga Halo Finder\footnote{\url{http://popia.ft.uam.es/AHF}} \citep[{\sc AHF};][]{AHF}. The distribution of $M_{500}$ masses\footnote{Masses and radii are expressed in terms of overdensities relative to the critical density of the Universe, at the cluster redshift: $M_{500} = (4\pi/3)\, 500\, \rho_\mathrm{c}(z)\, R_{500}^3$} (as reported in the {\sc AHF} catalogue) is shown in the left panel of Fig.~\ref{fig:sample}. In the middle panel, we show the distribution of the average (within $R_{500}$) projected spectroscopic-like temperature $T^{\rm sl}_{500}$ \citep{Mazzotta2004}, which mimics the X-ray spectroscopic temperature. The sample spans a mass range of $2$ to $26 \times 10^{14}{\rm M_\odot}$ (with a median of $7.7 \times 10^{14}{\rm M_\odot}$) and a temperature range between $2 \,\rm keV$ and $13 \,\rm keV$ (median: $5.4 \,\rm keV$).

\subsection{The dynamical state} \label{ssec:DS}

    Our analysis shares $266$ clusters with \citet[][hereafter \citetalias{DeLuca2021}]{DeLuca2021}, who also investigated their dynamical state. We therefore extended the dynamical and morphological analyses of \citetalias{DeLuca2021} (see Sect.~\ref{ssec:morph}) to the remaining clusters. In particular, we characterised the dynamical state by combining dynamical indicators in the relaxation indicator $\chi$ as suggested in \citet{Haggar2020}, but following the definition of \citetalias{DeLuca2021}
    \begin{equation}
        \chi = \left[\frac{1}{N}\sum_i^N \left(\frac{x_i}{x_{0,i}}\right)^2\right]^{-1/2},
        \label{eq:chi}
    \end{equation}
    where $x_i$ are the following dynamical indicators calculated within $R_{500}$: the total sub-halo mass fraction, $f_\mathrm{s}=\sum_i M_{i}/M_{500}$ \citep[the fraction of mass in sub-haloes relative to the total cluster mass; e.g.][]{Neto2007, Cui2017}, and the centre of mass offset with respect to the theoretical cluster centre: $\Delta_\mathrm{r}=|\mathbf{r}_{\rm cm}-\mathbf{r}_{\rm c}| /R_{500}$ \citep[e.g.][]{Crone1996, Maccio2007}, where the theoretical centre, $\mathbf{r}_{\rm c}$, is identified as the peak of the density field. As in \citetalias{DeLuca2021}, we set $x_{0,i}=0.1$ for the normalisation factors in Eq.~\eqref{eq:chi}. In general, the more relaxed a cluster is, the higher its value of $\chi$. This trend is more evident if we divide the sample with a traditional trichotomous classification. Following \citetalias{DeLuca2021}, we classified clusters as relaxed if $f_{\rm s}$ and $\Delta_{\rm r}$ are both lower than $0.1$ (and thus $\chi > 1$), disturbed if they are both greater than $0.1$ ($\chi < 1$), and hybrid otherwise. The right panel of Fig.~\ref{fig:sample} shows the $\chi$ distribution of the sample, while the three colours in all the plots refer to the three dynamical classes. The fractions of the relaxed ($\sim 57\%$), hybrid ($\sim 25\%$), and disturbed ($\sim 18\%$) systems are similar to those of \citetalias{DeLuca2021}.

\section{Method} \label{sec:meth}

    In \citetalias{Kozmanyan2019}, the distribution of the cluster structure parameter, $\mathcal{B}$, derived from hydrodynamical simulations, was used to extract cosmological information from observed discrepancies in temperature profiles measured by \citet{Bourdin2017}. In particular, \citet{Bourdin2017} combined X-ray and millimetre observations of $61$ galaxy clusters (with $0.047<z<0.447$) from \XMM{} and \Planck{} satellites, performing a joint deprojection of density, temperature, and pressure profiles with the parametric models of \citet{Vikhlinin2006a} and \citet{Nagai2007}, presented in the next section. \citetalias{Kozmanyan2019} adopted the same parametric ICM models and derived $\mathcal{B}$ values. More specifically, making explicit the profile normalisations in Eqs.~\eqref{eq:X_sb},~\eqref{eq:y}, for a given aperture $\theta$, we can write
    \begin{align}
        \Sigma_\mathrm{X}(\theta) &= \frac{n_{\mathrm{e},0}^2 C_\rho e_{\rm LOS}}{4\pi (1 + z)^4}\frac{n_\mathrm{p}}{n_\mathrm{e}}\left(1+4\frac{n_\mathrm{He}}{n_\mathrm{p}}\right) D_\mathrm{A} \theta f_\mathrm{X}, \label{eq:SX_eta} \\
        T_\mathrm{X}(\theta) &= T_0 f_\mathrm{T}, \label{eq:TX_eta} \\
        y(\theta) &= e_{\rm LOS} P_0 D_\mathrm{A} \theta f_{\rm SZ}, \label{eq:y_eta}
    \end{align}
    where $T_\mathrm{X}$ describes the projected spectroscopic-like temperature, $n_\mathrm{He}$ is the helium density, and $C_\rho$ and $e_{\rm LOS}$ are corrective factors accounting for deviations from the assumptions of gas homogeneity and spherical symmetry, such as gas clumpiness \citep[e.g.][]{Mathiesen1999, Eckert2015} and asphericity, respectively. The quantity $n_{\mathrm{e},0}$, $P_0$, and $T_0$ are the normalisations of the profiles with integrated LOS shapes $f_i$:
    \begin{align}
        & f_\mathrm{X} = \int \frac{n_\mathrm{e}^2/n_{\mathrm{e},0}^2}{\sqrt{x'^2 - x^2}}\Lambda_\mathrm{X} x' dx', \\
        & f_{\rm SZ} = \int \frac{P/P_0}{\sqrt{x'^2 - x^2}} x' dx', \\
        & f_\mathrm{T} = \frac{\int \frac{w T/T_0 x'}{\sqrt{x'^2 - x^2}} dx'}{\int \frac{wx'}{\sqrt{x'^2-x^2}}dx'},
    \end{align}
    where $w$ is the spectroscopic-like weight \citep[see next section;][]{Mazzotta2004}. Following \citetalias{Kozmanyan2019}, any systematic deviation between the normalisations of the temperatures can be expressed as
    \begin{equation}
        \eta_T = \frac{kT_0}{P_0/n_{e,0}} = b_n \cdot \mathcal{C}(z,\vec{\vartheta}) \cdot \mathcal{B},
        \label{eq:eta_biases}
    \end{equation}
    where $\vec{\vartheta}$ is the set of cosmological parameters and $z$ is the cluster redshift. Compared to \citetalias{Kozmanyan2019}, we generalise Eq.~\eqref{eq:eta_biases} by introducing an additional bias factor, $b_n$, to account for observational or signal-modelling systematics that are not included in $\mathcal{B}$ or $\mathcal{C}$, such as uncertainties in the X-ray temperature calibration or missing relativistic correction in the SZ effect. If clumpiness and asphericity are the only source of systematics, then $\mathcal{B} \propto \sqrt{C_\rho/e_{\rm LOS}}$ \citepalias[for a more detailed derivation of $\eta_T$, as well as $\mathcal{B}$ or $\mathcal{C}$, see Appendix A of][]{Kozmanyan2019}.
    
    Regarding $\mathcal{C}$, if our assumptions accurately describe clusters and their distances, then $\mathcal{C} = 1$. Otherwise, the underlying cosmological parameters should differ from the reference values adopted in the thermodynamic analysis. By combining Eqs.\eqref{eq:SX_eta},~\eqref{eq:TX_eta},~\eqref{eq:y_eta}, and~\eqref{eq:eta_biases}, we can express $\mathcal{C}$ as in \citetalias{Kozmanyan2019}
     \begin{align}
        \mathcal{C}(z, \vec{\vartheta}) &= \mathcal{D_A}(z, H_0,\{\Omega_i\}) \mathcal{Y}(Y), \label{eq:cosm_biases} \\
        \mathcal{D_A}(z, H_0,\{\Omega_i\}) &= \left( \frac{\bar{D}_\mathrm{A}(z, \bar{H}_0,\{\bar{\Omega}_i\})}{D_\mathrm{A}(z, H_0,\{\Omega_i\})} \right)^{1/2}, \label{eq:Dfunc} \\
        \mathcal{Y}(Y) &= \left( \frac{2+4Y\xi}{2-Y+2\xi} \right)^{1/2} \left( \frac{2+4\bar{Y}\xi}{2-\bar{Y}+2\xi} \right)^{-1/2}, \label{eq:Yfunc}
    \end{align}
    with $Y$ and $\xi$ the abundances of helium (which should be close to the primordial one) and metals, respectively. Barred quantities denote the reference cosmological values, such as for $H_0$ and $\Omega_\mathrm{m}$, assuming a flat $\Lambda$ cold dark matter ($\Lambda$CDM) cosmology.

\subsection{Thermodynamic profiles} \label{ssec:profiles}

    To measure $\mathcal{B}$ from the simulation sample, we first needed to create a mock dataset of \TheTH{} {\sc gadget-x} clusters. Following \citetalias{Kozmanyan2019}, the projected quantities were computed by summing over particles within cylindrical shells along the LOS, with a base area $A_{\rm ring}= \pi(r^2_{\rm max} - r^2_{\rm min})$. In particular, we derived the ICM densities from a proxy of the X-ray emission measure ($EM = \int n_\mathrm{p} n_\mathrm{e} dl$), while the projected LOS temperatures were computed using spectroscopic-like weights
    \begin{align}
        &EM(r_{\rm min}, r_{\rm max}) \propto \frac{\sum m_i \rho_i}{A_{{\rm ring}[r_{\rm min}, r_{\rm max}]}/\pi} , \label{eq:EM} 
        \\
        &T_\mathrm{X}(r_{\rm min}, r_{\rm max})=\frac{\sum w_i T_i V_i}{\sum w_i V_i} , \label{eq:Tsl}
    \end{align}
    with $m_i$, $\rho_i$, and $T_i$ the mass, density, and temperature of the particles within the considered volume $V_i$, respectively. For the Compton $y$, we considered a discretised version of Eq.~\eqref{eq:y}
    \begin{equation}
        y(r_{\rm min}, r_{\rm max})=\frac{\sigma_T}{m_\mathrm{e} c^2} \frac{\sum P_i V_i W}{A_{{\rm ring}[r_{\rm min}, r_{\rm max}]}},
        \label{eq:y_sim}
    \end{equation}
    where $P_i$ is the gas pressure and $W$ is the SPH smoothing kernel. For the integration length, we adopted a value of $20 \,\rm Mpc$ in this work, but we tested values of $10 \,\rm Mpc$ and $2R_{100}$ in Appendix~\ref{sec:los} to assess the effect of possible contaminants in the signal projection (both of physical and numerical origin, such as the presence of cosmic filaments or of low-resolution contaminant particles). In general, the three LOS lengths give consistent $\mathcal{B}$ distributions.
    
    The density, temperature, and pressure were then fitted from these projected quantities considering a forward approach \citep{Ettori2013} and the \citet{Vikhlinin2006a} and \citet{Nagai2007} profiles
    \begin{align}
        [n_\mathrm{p} n_\mathrm{e}](r) = &\, \frac{n_0^2(r/r_\mathrm{c})^{-\alpha}}{\left[1+(r/r_\mathrm{s})^\gamma\right]^{\epsilon/\gamma} \left[1+(r/r_\mathrm{c})^2\right]^{3\beta_1-\alpha/2}} + \label{eq:3D_Xprofne}
        \\
        & + \frac{n_{0,2}^2}{\left[1+(r/r_{\mathrm{c},2})^2\right]^{3\beta_2}}, 
        \nonumber
        \\
        T(r)= &\, T_0 \frac{x+T_{\rm min}/T_0}{x+1} \frac{(r/r_\mathrm{t})^{-a}}{\left[1+(r/r_\mathrm{t})^b\right]^{c/b}}, \label{eq:3D_Xprofte}
        \\
        \frac{P_\mathrm{e}(r)}{P_{500}} = &\, \frac{P_0}{(r/r_\mathrm{s})^\gamma \left[1+(r/r_\mathrm{s})^\alpha\right]^{(\beta-\gamma)/\alpha}}. \label{eq:3D_yprof}
    \end{align}
    As in \citet{Bourdin2017} or \citet{DeLuca2026}, we first jointly fitted the density and temperature templates to the mock X-ray proxies. In a second step, the SZ Compton parameter was used to derive the pressure and, in combination with the density, to build the radial temperature template $kT_\mathrm{e}^{\rm SZ,X}(r) = P^{\rm SZ}_\mathrm{e}(r)/n^\mathrm{X}_\mathrm{e}(r)$. This template was then projected with spectroscopic-like weights
    \begin{equation}
        T_{\rm SZ,X} = \eta_T \cdot T_{\rm sl}^{\rm SZ,X} = \eta_T \frac{\int w T_\mathrm{e}^{\rm SZ,X} dl}{\int w \, dl} .
        \label{eq:eta_T}
    \end{equation}
    Although Eq.~\eqref{eq:eta_T} is written in terms of projected spectroscopic-like temperature profiles, $\eta_T$ is here the same quantity defined in Eq.~\eqref{eq:eta_biases}, but measured considering the projected temperatures. Therefore, $\eta_T$ can be constrained by comparing Eq.~\eqref{eq:eta_T} with the spectroscopic-like temperature inferred from projected data, Eq.~\eqref{eq:Tsl}. In a simulation, the cosmological framework is known and hence $\mathcal{C} = 1$. In addition, no observational or signal modelling bias is included in the reconstruction of the thermodynamic profiles, and thus $b_n = 1$. Therefore, measuring $\eta_T$ in simulations directly provides $\mathcal{B} \equiv \eta_T = T_\mathrm{X}/T^{\rm SZ,X}_{\rm sl}$. In this way, $\eta_T$ quantifies the mismatch deviations introduced when the thermodynamic structure of a cluster is reconstructed using simplified models based on smooth, spherically symmetric ICM profiles, while the underlying cluster exhibits a more complex structure. This includes the effects of cluster asphericity, gas inhomogeneities, and possible mismatches associated with the different functional forms of the \citet{Vikhlinin2006a} and \citet{Nagai2007} models.
    
    We note that, for the derivation of $\mathcal{B}$, we are not interested in estimating the 3D properties of clusters but rather in their projected thermodynamic profiles. Thus, different projections from sufficiently separated LOS can be considered as ‘independent’ objects. We considered three orthogonal projections per cluster (along the coordinate system of the simulation cubes), and thus increased the sample size to $972$ objects. To test whether repeated haloes introduced residual biases in the results, we performed a Monte Carlo resampling in which only one projection per cluster was selected. For each realisation, we compared the $\mathcal{B}$ resampled distribution with the overall sample with a Kolmogorov-Smirnov (KS) two-tail test. With $10^6$ realisations, we found no significant biases, as presented in Appendix~\ref{sec:Bvsproj}.

\subsection{The morphological state} \label{ssec:morph}

    As suggested observationally by \citet{DeLuca2026} for the CHEX-MATE sample \citep{CHEX-MATE}, the dynamical and morphological states should affect the distribution of $\mathcal{B}$. To study this relation, we extended the morphological analysis of \citetalias{DeLuca2021} and \citet{Campitiello2022}. In particular, mock Compton-$y$ and X-ray maps were generated with the {\sc pymsz}\footnote{\url{https://github.com/weiguangcui/pymsz}} \citep{Cui2018} and {\sc pyxsim} \citep{Biffi2012, Zuhone2014, pyXSIM} packages. The maps covered a region of radius $1.4R_{200}$ around the (projected) highest peak of the 3D density distribution, and were generated with the same spatial resolution of $10 \,{\rm kpc/pixel}$, without including contaminants or instrumental noise \citepalias[for further details, see][]{DeLuca2021}.

    Similarly to the dynamical state parameter, we characterised the morphology using the combined parameter $\mathcal{M}$ \citep{rasia2013, Cialone2018, DeLuca2021}. It is defined as a weighted sum of standardised morphological indicators
    \begin{equation}
        \mathcal{M}=\frac{1}{\sum_i W_i}\left(\sum_i W_i \dfrac{\log(V_i^{\alpha_i})-<\log(V_i^{\alpha_i})>}{\sigma_{\log(V_i^{\alpha_i})}}\right) ,
        \label{eq:M}
    \end{equation}
    where $V_i$ is a set of morphological parameters, $W_i$ their associated weights, and $\sigma_{\log(V_i)}$ their standard deviation. The parameter $\mathcal{M}$ provides a relative ranking of the cluster relaxation within the sample. The more negative $\mathcal{M}$, the more relaxed the cluster is, and the opposite is true for disturbed systems. For the set of morphological parameters, both the classification and the definition of the parameters generally vary from study to study, such as in the choice of centres or apertures. Since we are interested in studying the relation between morphology and $\mathcal{B}$, and to use it as an informative prior for future cosmological analysis, we replicated the morphological analysis done for the observed CHEX-MATE sample presented in \citet{Campitiello2022}. Similarly to their work, we smoothed the simulated X-ray images to mimic \XMM{} observations and defined the $\mathcal{M}_\mathrm{X}$ indicator as the simple average (i.e. $W_i=1$) of the light concentration ratio $c$ \citep{Santos2008}, the centroid shift $w$ \citep{Mohr1993}, and the third ($P_3/P_0$) and second order ($P_2/P_0$) power ratios \citep{Buote1995}. The apertures for the parameters were set to $R_{500}$ (and $0.15R_{500}$ for the inner one of $c$). In addition, we expanded the morphological analysis of \citetalias{DeLuca2021} in Appendix~\ref{sec:Morp_300}. In particular, we considered $c$, $w$, the asymmetry $A$ \citep{Schade1995}, $P_3/P_0$, and the Gaussian fit ($G$) and strip ($S$) parameters \citep{Cialone2018}, adopting the same optimised apertures and weights of \citetalias{DeLuca2021} for SZ ($\mathcal{M}_{\rm SZ}$) and X-ray ($\mathcal{M}^{\rm DL21}_\mathrm{X}$) maps.

\subsection{Informative priors for cosmological inferences} \label{ssec:BtoH0}

    In \citetalias{Kozmanyan2019}, the cosmological inference was corrected for the cluster structure bias by considering the distribution of $\mathcal{B}$ from simulations as an informative prior in the Bayesian analysis. More precisely, for a sample of $N$ clusters, they introduced $N$ nuisance parameters $\mathcal{B}_i$, each following the same distribution suggested from the simulations. The posterior distribution of the cosmological parameter of interest, $\vartheta_k$, was then calculated by marginalising over
    \begin{equation}
        P(\vartheta_k) \propto \int \mathcal{L}\left(\vec{\eta_T}\big\rvert \vec{\vartheta}, \vec{\mathcal{B}}\right) p\left(\vec{\mathcal{B}}\right) p(\vec{\vartheta}) \, d\vec{\mathcal{B}} d\vec{\vartheta}_{-k},
        \label{eq:H0_likelihood}
    \end{equation}
    where $\vec{\vartheta}_{-k} = \{\vartheta_{i\neq k}\}$, and $\mathcal{L}$ is the likelihood function. If the measurements are not correlated (as expected for $\eta_T$ measured from $N$ different clusters), we can rewrite Eq.~\eqref{eq:H0_likelihood} as the product of the individual cluster likelihoods: $\mathcal{L}(\vec{\eta_T}\big\rvert \vec{\vartheta}, \vec{\mathcal{B}}) p(\vec{\mathcal{B}}) = \prod_i \mathcal{L}_i(\eta_T^{(i)}\big\rvert \vec{\vartheta}, \mathcal{B}_i) p(\mathcal{B}_i)$. In \citetalias{Kozmanyan2019}, the marginalisation was performed numerically with a Markov chain Monte Carlo (MCMC) algorithm, assuming a flat $\Lambda$CDM cosmology. In this work, we propose to refine this approach by introducing an explicit dependence on the cluster dynamical state. In fact, the distribution of $\mathcal{B}$ is expected to be affected by the cluster relaxation, as observed by \citet{DeLuca2026} for the CHEX-MATE clusters. 
    
    To model this dependence, we examined two strategies. For the first approach, we considered a traditional trichotomous classification dividing the sample into relaxed, hybrid, and disturbed systems (the 3DS model). With this framework, class-based informative priors can be introduced in the cosmological analysis by dividing the observed sample in a similar way. Although this approach accounts for differences among dynamical states, it treats dynamical information as discrete, whereas cluster relaxation is intrinsically a continuous quantity \citep[e.g. \citetalias{DeLuca2021};][]{Haggar2024}. Moreover, in real observations the dynamical state cannot be measured directly using 3D indicators. Classifications are made from a morphological point of view, which may be prone to systematics such as projection effects.
    
    For this reason, we implemented a data-driven regression based on Gaussian processes \citep[$GPs$; see, e.g.][]{RasmussenGP} to model the relation between $\mathcal{B}$ and the X-ray combined morphological indicator, $\mathcal{M}_\mathrm{X}$. In $GP$ formalism, an unknown function is modelled as a stochastic process defined by a mean $\mu$ and a covariance $k$ function: $f(x) \sim GP[\mu(x),k(x,x')]$. Given these priors, $f(x)$ values are drawn from a multivariate Gaussian distribution, and values at new positions $x_*$ can be predicted by conditioning this multivariate distribution
    \begin{equation}
        \begin{bmatrix}
            f(x) \\
            f(x_*)
            \end{bmatrix}
            \sim \mathcal{N}\!\left(
            \begin{bmatrix}
            \mu(x) \\
            \mu(x_*)
            \end{bmatrix},
            \begin{bmatrix}
            k(x,x') & k(x_*,x) \\
            k(x_*,x) & k(x_*,x'_*)
            \end{bmatrix}
        \right).
    \end{equation}
    An advantage of $GPs$ is that the behaviour of $f(x)$ is fully determined by the choice of the mean and covariance functions (and thus by their hyperparameters). More specifically, the covariance (also referred to as kernel) controls the smoothness, characteristic length scales, and the amplitude of the variations. The mean instead sets the expected values of $f(x)$.
    
    In this work, we tested two different $GP$ set-ups to describe heteroscedastic data. As a first pragmatic attempt to incorporate non-constant intrinsic scatter, we implemented a $GP$ regression model based on the \texttt{scikit-learn} tools \citep{scikit-learn}. However, \texttt{scikit-learn} does not include a native implementation of $GP$ regression in which variance is treated as an explicit input-dependent latent function and inferred jointly with the mean. Thus, we adopted an approximate heuristic construction in which the kernel structure allows the variance to depend on the input data, and we refer to this model as $GP_{\rm kernel}$. In particular, we defined the $GP_{\rm kernel}$ model as the sum of three kernel components: $k(x,x') = k_{\rm RBF}(x,x') + k_{\rm WN}(x,x') + k_{\rm DP}(x,x')k_{\rm WN}(x,x')$, where $k_{\rm RBF}$ is the squared exponential kernel (also called the radial basis function)
    \begin{equation}
        k_{\rm RBF}(x,x';\ell)= a^2 \exp\left[-\frac{(x-x')^2}{2\ell^2}\right],
        \label{eq:exp_quad}
    \end{equation}
    with $a$ the amplitude and $\ell$ the characteristic correlation length scale. This kernel controls the smoothness and captures non-linear relations. We also introduced a white noise kernel ($k_{\rm WN}$) in the covariance, which is zero for $x \neq x'$, and is constant otherwise. Finally, the product term includes a linear dot-product kernel ($k_{\rm DP}$) that introduces a monotonic dependence of the variance on the input values (and thereby approximating a form of heteroscedasticity) by affecting only the diagonal of the covariance matrix: $k_{\rm WN}(x_i,x_j) + k_{\rm DP}(x_i,x_j) k_{\rm WN}(x_i,x_j) \approx [\sigma^2_0+\sigma^2 x_i\cdot x^j]\delta(x_i-x_j)$. To regularise the sampling density along the inputs and stabilise the kernel optimisation, we did not use $\mathcal{M}_\mathrm{X}$ directly. Instead, we mapped it onto a monotonic normalised variable: $x=\Phi(\frac{\mathcal{M}_\mathrm{X} - \langle \mathcal{M}_\mathrm{X} \rangle}{\sigma(\mathcal{M}_\mathrm{X})})$, where $\Phi$ is the cumulative of the normal distribution. This monotonic transformation also introduces a physically motivated trend in the dispersion, which is expected to increase from relaxed to disturbed systems. The model was then trained on this transformed input using a Gaussian likelihood. In this way, the model provides a simple and computationally efficient way to encode a monotonic trend in the scatter. However, it does not provide a full probabilistic model with learnt input-dependent likelihood inferred jointly with the mean function (e.g. it may not learn the underlying heteroscedastic trend).
    
    Therefore, for a statistically coherent description of both the mean and variance relations, we also implemented a fully probabilistic heteroscedastic model ($GP_{\rm hetero}$) with the \texttt{PyMC} package \citep{pymc2023}. Following the \texttt{PyMC} example\footnote{\url{https://www.pymc.io/projects/examples/en/latest/gaussian_processes/GP-Heteroskedastic.html}}, we modelled the mean and logarithmic standard deviation functions with two independent $GP$. For both components, we adopted squared exponential kernels with informative priors on their hyperparameters ($a \sim \mathrm{Gamma}$, $\ell \sim \mathrm{Inverse\,Gamma}$) to stabilise the inference (e.g. to disfavour implausible correlation length scales\footnote{\url{https://betanalpha.github.io/assets/case_studies/gp_part3/part3.html\#4_adding_an_informative_prior_for_the_length_scale}}). To reduce computational cost, we adopted a sparse $GP$ formulation based on inducing points selected via $k$-means on the input $\mathcal{M}_\mathrm{X}$ values. Inference was performed with a no-u-turn sampler \citep[NUTS;][]{Hoffman2014} with Gaussian likelihood. Hereafter, unless otherwise stated, chain convergence was assessed using the Gelman--Rubin diagnostic \citep{Gelman1992, Vehtari2021}, requiring $\hat{R}< 1.01$ for all parameters. In this way, the $GP$ prediction includes the marginalisation over the uncertainties in both latent functions (mean and intrinsic scatter) and their kernel hyperparameters. As a final consideration, we note that \citet{DeLuca2026} found that the posterior distributions of the $\eta_T$ values (and thus of $\mathcal{B}$) are generally asymmetric and well described by log-normal distributions (consistently with their definition as a product of positive independent terms; see Eq.~\ref{eq:eta_biases}). For this reason, the $GP$ regressions (with the assumed Gaussian likelihood) are performed on the $\ln \mathcal{B}$ values.

\section{Results} \label{sec:res}

\subsection{Distribution of $\mathcal{B}$ values} \label{ssec:Bdist}

    With the pipeline defined in Sect.~\ref{ssec:profiles} (integrating the profiles up to $20 \,\rm Mpc$, see Appendix~\ref{sec:los}), we show in Fig.~\ref{fig:Bdist} the $\mathcal{B}$ distribution of the full simulated sample, together with the distribution of the relaxed ($\mathcal{B}_{\rm rel}$), hybrid ($\mathcal{B}_{\rm hyb}$), and disturbed ($\mathcal{B}_{\rm dis}$) subsamples. In general, the overall distribution is similar to that presented in \citetalias{Kozmanyan2019}, as shown in the inset panel of Fig.~\ref{fig:Bdist}. It is positively skewed ($\gamma = 1.18$) and leptokurtic (excess kurtosis: $\kappa = 4.2$), mainly because $\mathcal{B}$ is strictly positive and due to the dynamical state of the clusters. In fact, disturbed clusters generally have higher values and a larger dispersion than relaxed clusters, which are closer to $\mathcal{B} = 1$. The hybrid systems instead show an intermediate behaviour, with values close to unity but with a larger spread compared to relaxed clusters. A similar trend was also confirmed by \citet{DeLuca2026} for the observed clusters.

    \begin{figure}
        \centering
        \includegraphics[width=\columnwidth]{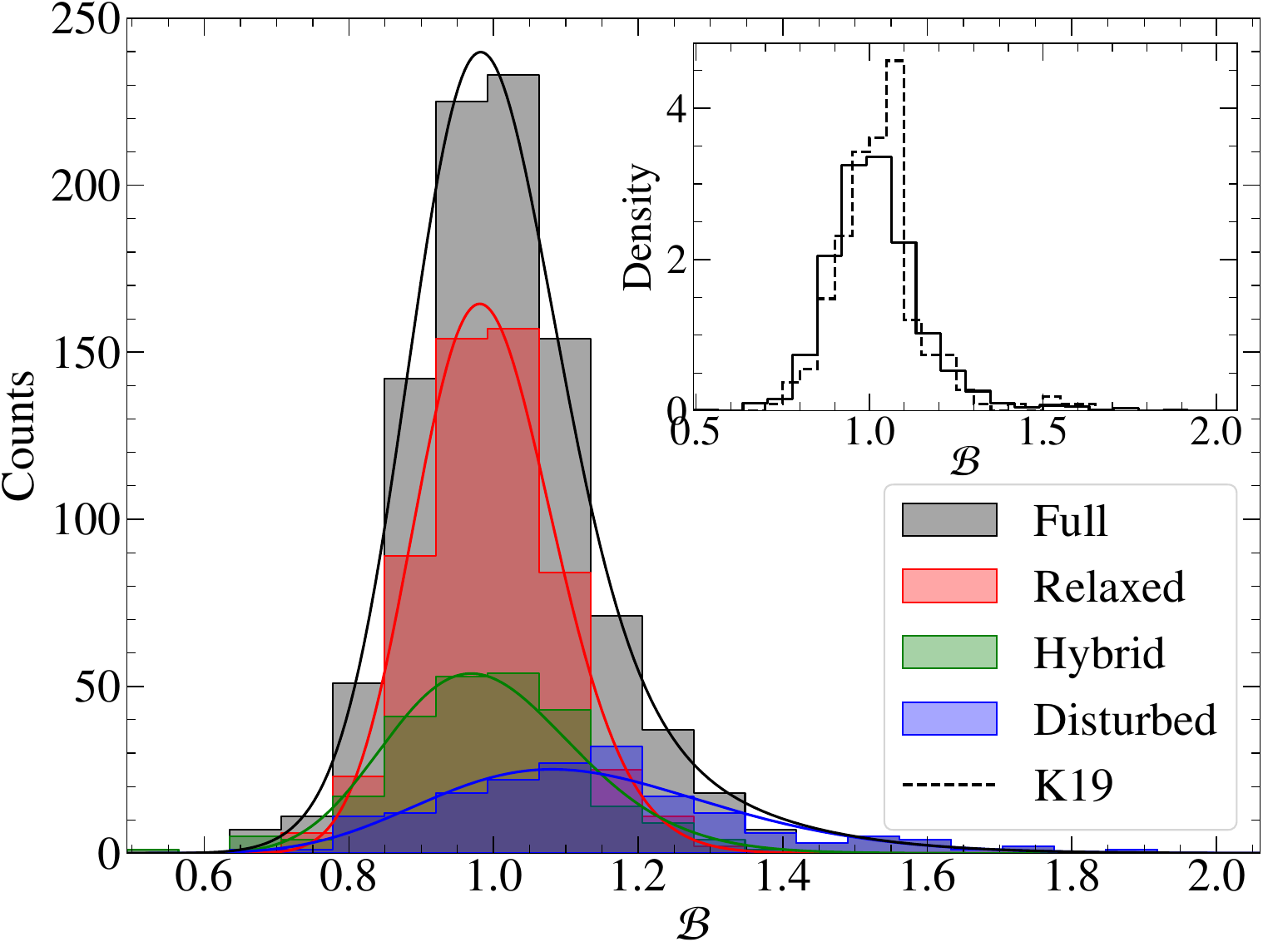} 
        \caption{Distributions of $\mathcal{B}$ for the full sample (grey histogram) and the relaxed (red), hybrid (green), and disturbed (blue) subsamples, together with their best fitting log-normal models (solid curves). The black line shows the combined 3DS class-based model, obtained from the three log-normal components fitted to the dynamical subsamples. A comparison between the \citetalias{Kozmanyan2019} distribution and our findings is shown in the inset panel with a dashed line.}        
        \label{fig:Bdist}
    \end{figure}

    Therefore, although the overall distribution does not show clear signs of bimodality (or more complex features), a single distribution may not describe the entire dataset. For this reason, we compared the results obtained by fitting a single overall distribution with a class-conditional model, based on the three dynamical classes (the 3DS model). In particular, we found that, among the most common skewed \texttt{scipy} distributions (tested with the \texttt{fitter}\footnote{\url{https://fitter.readthedocs.io/en/latest/}} package), log-normal distributions generally provide the best description in terms of the Bayesian information criterion (BIC). In Table~\ref{Tab:B20_stat}, we compare the single log-normal and class-conditional models, made of three log-normal components (shown in Fig.~\ref{fig:Bdist} with the black curve). For parameter estimates, we adopted weakly informative priors: $\mu \sim \mathcal{U}(-1,1)$ and $\ln(\sigma)\sim \mathcal{U}(-6,0)$. Posterior samples were drawn using the \texttt{PyMC} NUTS sampler, with 12 chains of 5000 iterations each after 1000 tuning steps. Model comparison was performed using the BIC and the maximum log-likelihood over posterior samples. Compared to the single log-normal model, the 3DS model yields a lower BIC, $\Delta\mathrm{BIC}=222$, and a higher maximum log-likelihood: $\Delta\ln\hat{\mathcal{L}}=125$. Therefore, the class-conditional model is favoured, despite its larger number of parameters.

    \begin{table}
        \caption{Summary statistics of the $\mathcal{B}$ distributions.}
        \centering
        \setlength{\tabcolsep}{3pt}
        \begin{tabular}{ c  c  c  c  c  c }
            \hline\hline
            Sample & Size & Median & IQR & $\mu_{\rm fit} [\times10^{-2}]$ & $\sigma_{\rm fit} [\times10^{-2}]$ \\ 
            \hline 
            $\mathcal{B}$ & \multirow{2}{*}{972} & \multirow{2}{*}{1.01} & \multirow{2}{*}{0.15} & \multirow{2}{*}{$1.08 \pm 0.43$} & \multirow{2}{*}{$13.29 \pm 0.30$} \\ 
             (Single $\mathcal{LN}$) & & & & \\
            $\mathcal{B}_{\rm rel}$ & 552 & 0.99 & 0.10 & $-0.98 \pm 0.41$ & $9.68 \pm 0.29$  \\ 
            $\mathcal{B}_{\rm hyb}$ & 246 & 0.99 & 0.13 & $-1.38 \pm 0.85$ & $13.28 \pm 0.60$ \\ 
            $\mathcal{B}_{\rm dis}$ & 175 & 1.12 & 0.21 & $11.04 \pm 1.37$ & $18.00 \pm 0.97$ \\ 
            \hline \hline
             & \multicolumn{3}{c}{Single $\mathcal{LN}$ model} & \multicolumn{2}{c}{3DS model} \\
             \hline
              $\ln{\hat{\mathcal{L}}}$ & \multicolumn{3}{c}{573} & \multicolumn{2}{c}{698} \\
              $BIC$ & \multicolumn{3}{c}{-1133} & \multicolumn{2}{c}{-1355} \\
            \hline \hline            
        \end{tabular} 
        \tablefoot{The table reports the sample size, median, interquartile range (IQR), and fit results. The parameters of the log-normal models are given with their $68\%$ confidence intervals.}
        \label{Tab:B20_stat}
    \end{table}

\subsection{Correlation with cluster properties} \label{ssec:BvsDMS}

    \begin{figure}
        \centering
        \includegraphics[width=\columnwidth]{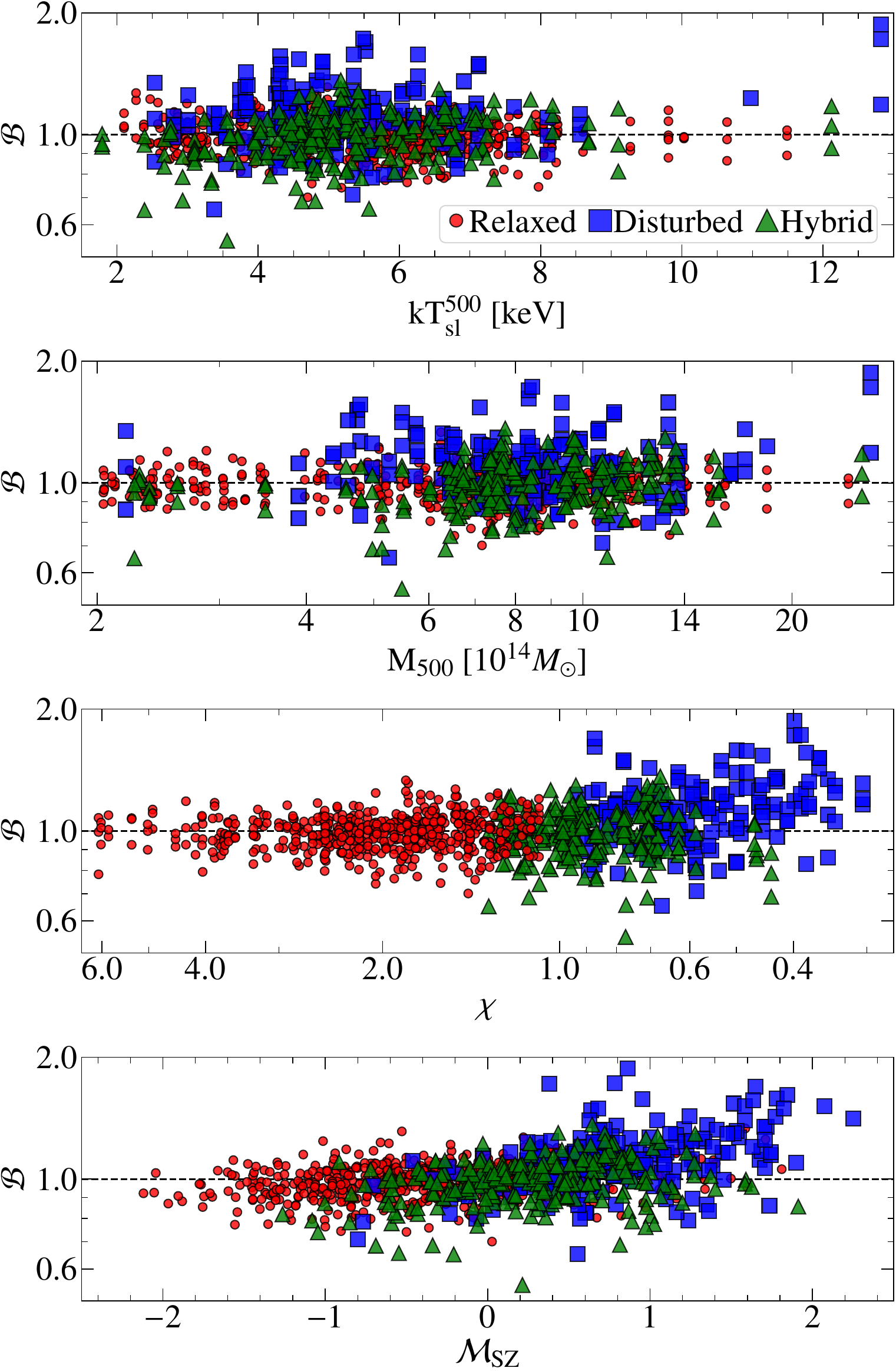} 
        \caption{Correlations between $\mathcal{B}$ and the cluster temperature, mass, and the dynamical ($\chi$) and morphological ($\mathcal{M}_{\rm SZ}$, for SZ maps) indicators, respectively from upper to lower panels. Relaxed clusters are shown as red circles, while green triangles and blue squares are used for hybrid and disturbed systems. The $\chi$ axis is reversed to match the relaxation ordering of $\mathcal{M}_{\rm SZ}$.}        
        \label{fig:Bcors}
    \end{figure}

    Given the dependence of the $\mathcal{B}$ distribution on the dynamical classification, we investigated possible correlations between $\mathcal{B}$ and dynamical and morphological indicators, as well as other cluster properties such as temperatures and masses. In Figs.~\ref{fig:Bcors} and~\ref{fig:GPres}, we show the distribution of $\mathcal{B}$ as a function of these quantities. We find no significant correlations with spectroscopic temperatures (Spearman $p\text{-value}\sim 1$) or masses ($\rho \sim 0.06$, $p \sim 0.06$). With continuous indicators to trace the dynamical and morphological states, we can draw similar conclusions to the class-based analysis of Sect.~\ref{ssec:Bdist}. In particular, we observe an increase in the scatter of $\mathcal{B}$ values from the most relaxed (e.g. those with the lower values of $\mathcal{M}$ or the higher $\chi$) to the most disturbed clusters. Moreover, $\mathcal{B}$ weakly correlates with $\chi$ ($\rho = -0.23$) and $\mathcal{M}$ ($\rho = 0.38$ for SZ maps, $\rho = 0.33$ for X-ray).

\subsection{Gaussian process results} \label{ssec:GP}

    \begin{figure}
        \centering
        \includegraphics[width=\columnwidth]{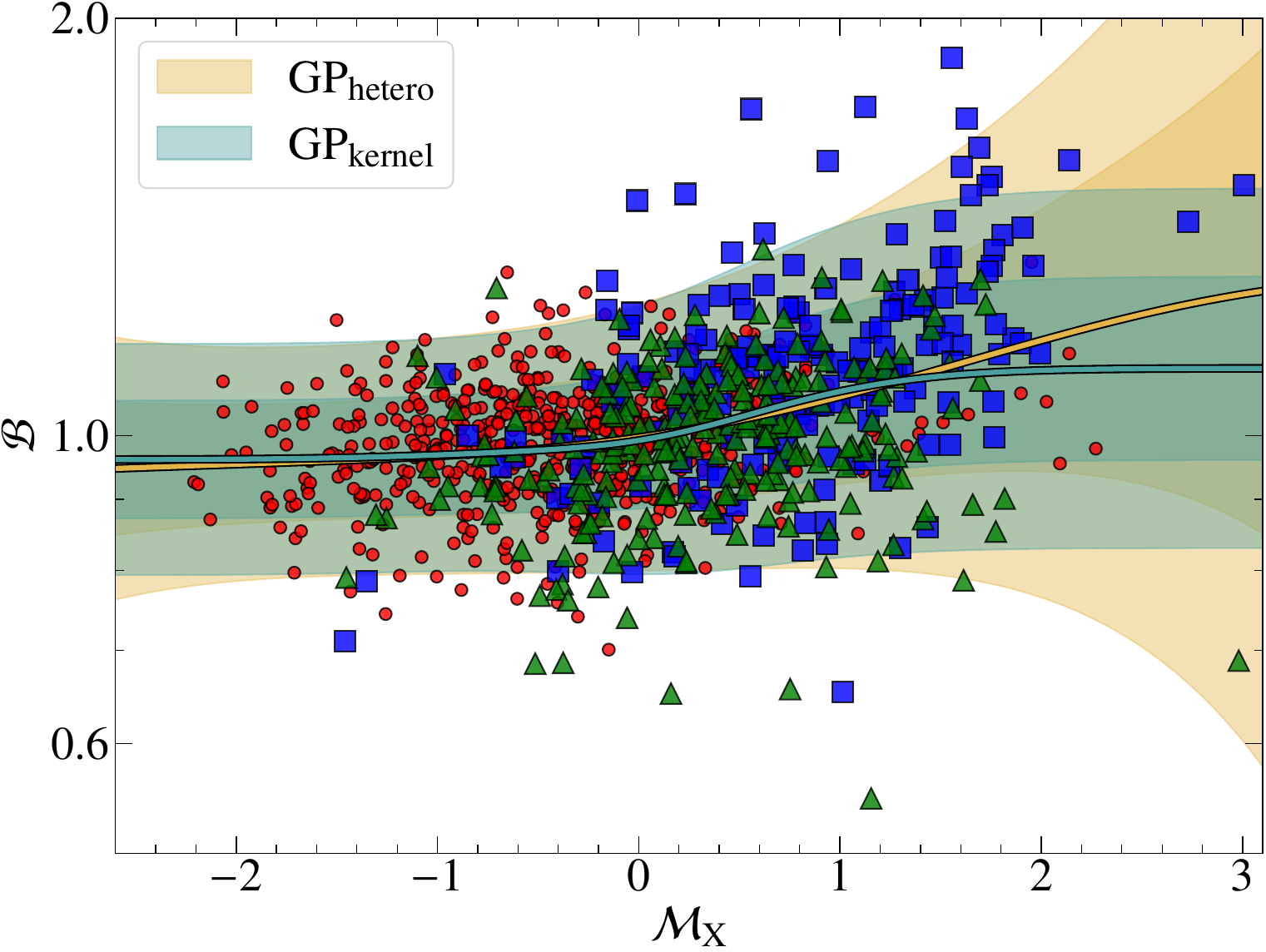} 
        \caption{Results of the $GP$ regressions with $1\sigma$ and $2\sigma$ dispersions for the $GP_{\rm kernel}$ (muted teal) and $GP_{\rm hetero}$ (muted amber) models. The trichotomous classification is shown with red circles for relaxed, green triangles for hybrid, and blue squares for disturbed systems, respectively.}
        \label{fig:GPres}
    \end{figure}
    
    As discussed in Sect.~\ref{ssec:BtoH0}, the trends with the dynamical and morphological states of the clusters are expected and can be used to deliver informative priors for cosmological analyses. In Fig.~\ref{fig:GPres}, we show the results of the $GP$ regressions described in Sect.~\ref{ssec:BtoH0}. Within the explored range of $\mathcal{M}_\mathrm{X}$ values, the two models provide consistent descriptions. Instead, differences are present towards the tails of the $\mathcal{M}_\mathrm{X}$ distribution, reflecting the different $GPs$ modelling. The full heteroscedastic $GP_{\rm hetero}$ model is designed to learn both the underlying mean relation and the intrinsic scatter directly from the $(\ln\mathcal{B}$, $\mathcal{M}_\mathrm{X}$) data. Therefore, in underpopulated regions or beyond the sampled values, its prediction is characterised by an increased uncertainty due to the extrapolation process.
    
    In the $GP_{\rm kernel}$ model, instead, regression is performed after mapping the input to a different space, based on the cumulative distribution of the morphological indicator (see Sect.~\ref{ssec:BtoH0}), which is by construction limited to the interval $[0, 1]$. This transformation compresses the extremes of the $\mathcal{M}_\mathrm{X}$ distribution and, thus, limits the variation of the inferred behaviour in the tails. For clusters more relaxed (or disturbed) than those probed by our sample, this model provides similar behaviours to those at the tails of the observed distribution. As a consequence, the informative priors for the cosmological pipeline provided by the two $GPs$ may differ. More specifically, with the $GP_{\rm kernel}$ model, clusters more relaxed than those observed in this work (e.g. with $\mathcal{M}_\mathrm{X} \lesssim -2.5$), which should be closer to the spherical shape assumption, are described by a similar mean and covariance of the nearest sampled values. Therefore, the $GP_{\rm kernel}$ model will deliver a less conservative prior than the $GP_{\rm hetero}$ model.

\subsection{Cosmological pipeline validation and calibration} \label{ssec:H0_valid}
    
    With the trichotomous class-conditional model and the $GPs$ results, we studied the reliability and limitations of the cosmological pipelines based on simulated informative priors, as described in Sect.~\ref{ssec:BtoH0}. For these tests, we generated mock $\eta^{\rm mock}_T$ catalogues as follows. Given a sample size $N$, we randomly extracted $\mathcal{B}$ values from the simulated sample. To mimic an observation, we used Eq.~\eqref{eq:eta_biases} assuming for the underlying ‘true’ cosmology a $\Lambda$CDM model with $h_{\rm ref}=0.74$, and the \citet{Planck2020VI} values for matter density, $\Omega^{\rm ref}_m = 0.3153$, and a helium abundance of $Y^{\rm ref} = 0.242$, unless otherwise stated. Finally, we assigned to each cluster an uncertainty of $14\%$ and a random redshift between $0.05$ and $0.6$, approximately spanning the redshift interval of the CHEX-MATE observed sample \citep{CHEX-MATE}
    \begin{equation}
        \ln \eta^{\rm mock}_T = \ln\left[C(z;H_0^{\rm ref}, \Omega_\mathrm{m}^{\rm ref}, Y^{\rm ref})\right] + \ln\mathcal{B} + \mathcal{N}(0, 0.14),
        \label{eq:eta_mock}
    \end{equation}
    with an additional log-normal dispersion (of the same order) to simulate real observations consistent with the observational findings of \citet{DeLuca2026}. With these mock catalogues, we then tested our Bayesian analysis with the \texttt{PyMC} package, adopting the prior definitions summarised in Table~\ref{tab:pr_Setup}, unless otherwise specified. For the cosmological parameters, we fitted a flat $\Lambda$CDM model, with a uniform prior on $H_0$ and truncated normals for $\Omega_\mathrm{m}$ and $Y$ (between zero and one), based on the \citet{Planck2020VI} results. 
    
    \begin{table}
        \centering
        \caption{Common prior set-up for the cosmological inference tests.}
        \begin{tabular}{ c c c }
            \hline\hline
            \multicolumn{3}{c}{Cosmological priors} \\
            \hline 
            $h$ & \multicolumn{2}{c}{$\mathcal{U}(0.4, 1)$} \\
            $\Omega_\mathrm{m}$ & \multicolumn{2}{c}{$\mathcal{TN}(0.3153, 0.0073, 0, 1)$}  \\
            $Y$ & \multicolumn{2}{c}{$\mathcal{TN}(0.242, 0.012, 0, 1)$} \\
            \hline\hline
            \multicolumn{3}{c}{Cluster priors} \\
            \hline
            \multirow{3}{*}{3DS model} & $\mu_{i}$ & $\mathcal{N}(\mu_i^{fit}, \sigma^{fit}_i)$ \\
             & $\ln(\sigma_i)$ & $\mathcal{U}(-4,-0.7)$ \\
             & $\mathcal{B}_i$ & $\mathcal{LN}(\mu_i, \sigma_i)$ \\
            $GP$ models & $\mathcal{B}(\mathcal{M}_X)$ & $\mathcal{LN}[\mu_{GP}(\mathcal{M}_X), \sigma_{GP}(\mathcal{M}_X)]$ \\
            \hline\hline
        \end{tabular}
        \tablefoot{Here, $\mathcal{TN}(\mu,\sigma,a,b)$ denotes a normal distribution with mean $\mu$ and standard deviation $\sigma$, truncated between the lower and upper bounds $a$ and $b$.}
        \label{tab:pr_Setup}
    \end{table}
    
    For the informative prior on cluster structure bias, we note that both the trichotomous classification and the $GP$ models describe the $\mathcal{B}$ values with log-normal distributions: $\ln \mathcal{B}\sim\mathcal{N}(\mu_B,\sigma_B^2)$. Therefore, in log-space and assuming a Gaussian likelihood, Eq.~\eqref{eq:H0_likelihood}  can be simplified and the likelihood analytically marginalised over $\ln \mathcal{B}$
    \begin{equation}
    \begin{aligned}
        \mathcal{L}\left(\ln \vec{\eta_T}\big\rvert \vec{\vartheta}, \vec{\phi}\right) &= \int \mathcal{L}\left(\ln \vec{\eta_T}\big\rvert \vec{\vartheta}, \ln \vec{\mathcal{B}}\right) p\left(\ln \vec{\mathcal{B}}\right) d\ln \vec{\mathcal{B}} \\
        &= \mathcal{N}\left[\ln C\left(\vec{z};\vec{\vartheta}\right) + \vec{\mu}_B,\; \vec{\sigma}^{\,2}_{\ln \eta_T} + \vec{\sigma}^{\,2}_B \right] ,
        \label{eq:marglike}
    \end{aligned}
    \end{equation}
    where $\vec{\phi} = [\vec{\mu}_B, \vec{\sigma}_B]$ are the informative prior hyperparameters. Therefore, the marginal likelihood for the cosmological parameters remains Gaussian. Eq.~\eqref{eq:marglike} directly describes the marginal likelihood for the $GP$ models, with $\vec{\mu}_B\equiv\vec{\mu}_{GP}$ and $\vec{\sigma}_B\equiv\vec{\sigma}_{GP}$ given by the predicted mean and scatter of the $GP$ regressions.
    
    For the trichotomous classification, we implemented a class-based hierarchical inference. By dividing the sample into relaxed, hybrid, and disturbed systems, with the same class fractions as in the original sample, the total likelihood can then be written as the product of three marginal likelihood terms of Eq.~\eqref{eq:marglike}
    \begin{equation}
        \mathcal{L}\left(\ln\vec{\eta_T}\big\rvert \vec{\vartheta}, \vec{\phi_{\rm rel}}, \vec{\phi_{\rm hyb}}, \vec{\phi_{\rm dis}}\right) = \prod_{i\in cl} \mathcal{L}\left(\ln \vec{\eta}_{T,i}\big\rvert \vec{\vartheta}, \vec{\phi_i}\right),
        \label{eq:marg_like_3c}
    \end{equation}
    where $\vec{\phi_i} = [\mu_i, \sigma_i]$ are the set of informative prior parameters defined by the fitted log-normal distributions of the $i$th dynamical subsample, described in Sections~\ref{ssec:BtoH0} and~\ref{ssec:Bdist} (Table~\ref{Tab:B20_stat}). In this way, the cosmological parameters are jointly constrained by all clusters, while the information on their structure is injected as informative priors in the models differently, depending on their dynamical state. In particular, for the trichotomous class-based model we adopted Gaussian priors on $\mu_i$, while for $\sigma_i$ we assigned a scale invariant prior to give more flexibility to the model in case of possible mismatches between the simulation and the (future) observed data (see Table~\ref{tab:pr_Setup}).

\subsubsection{Prior informativeness and parameter degeneracies} \label{sssec:priors}

    \begin{figure}
        \centering
        \includegraphics[width=\columnwidth]{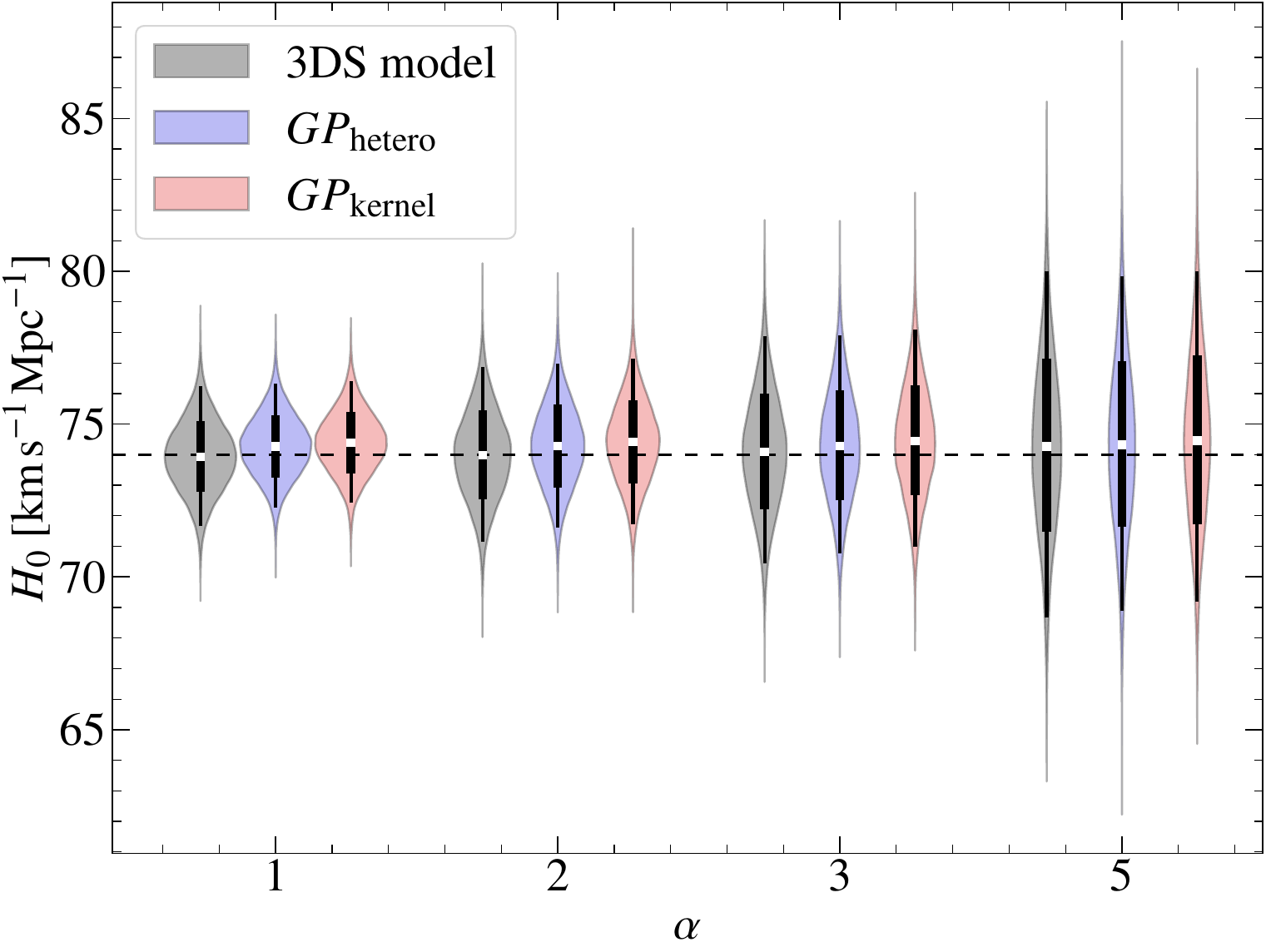} 
        \caption{Posterior distributions of $H_0$ for different prior strengths on $\Omega_\mathrm{m}$ and $Y$. The violin shapes show the marginal posterior densities and are symmetric by construction. White markers show the medians, while thick and thin bars represent the $68\%$ and $95\%$ percentile intervals, respectively. The dashed line marks the reference $H_0$.}
        \label{fig:Pr_violin}
    \end{figure}

    To quantify the sensitivity to the cosmological priors and the constraining power of the method, we performed four MCMC runs in which the priors in $\Omega_\mathrm{m}$ and $Y$, reported in Table.~\ref{tab:pr_Setup}, were progressively relaxed. In particular, we increased the standard deviation of these priors by a factor $\alpha = (1,2,3,5)$. The test was performed for a sample of 972 clusters using ten chains of 5000 iterations (plus 1000 of tuning) each with the \texttt{PyMC} slice sampler \citep{Neal2003}. The resulting $H_0$ posteriors are summarised in Fig.~\ref{fig:Pr_violin}, while the marginal posteriors of the other cosmological parameters are shown in Appendix~\ref{sec:Priors}, Fig.~\ref{fig:Pr_corner}. Independently of the method, $H_0$ remain constrained by the data, although the uncertainties increase for higher $\alpha$.
    
    The posterior distributions of $\Omega_\mathrm{m}$ and $Y$ are instead essentially prior-dominated for all $\alpha$, with the prior-to-posterior ratio of the standard deviations always close to unity. In a flat $\Lambda$CDM framework, $\eta_T$ depends mainly on three parameters: $H_0$, $\Omega_\mathrm{m}$, and $Y$, as described by Eq.~\eqref{eq:cosm_biases}. However, only one observable is fitted and this cosmological probe is intrinsically less sensitive to distances than other methods \citep[$\eta_T \propto D_\mathrm{A}(z)^{-1/2}$; e.g.][]{Allen2011}. Degeneracies among the cosmological parameters are therefore expected (see Appendix~\ref{sec:Priors}). At fixed $\Omega_\mathrm{m}$, keeping $\mathcal{C}$ constant gives approximately $H_0(Y)/H_0^{\rm ref} \simeq (2-Y)/(2-Y_{\rm ref})$, for $\xi\ll1$. In contrast, $\Omega_\mathrm{m}$ mainly affects the shape of $D_\mathrm{A}(z)$, and hence of $\eta_T(z)$. As a consequence, its degeneracy with $H_0$ depends on the redshift distribution of the sample. Thus, constraining $\Omega_\mathrm{m}$ would require small uncertainties and adequate redshift sampling. For these reasons, this method is primarily an $H_0$ probe, provided that reasonable priors are adopted for the remaining parameters, as done in \citetalias{Kozmanyan2019}; \citet{Ettori2020, Wan2021}. A more detailed discussion of these degeneracies is given in Appendix~\ref{sec:Priors}, where we show that variations of about $5\%$ in the central values of the $\Omega_\mathrm{m}$ and $Y$ priors generally induce shifts of less than $1\%$ in the inferred $H_0$. With this in mind, we focus primarily on $H_0$ in the following tests.

\subsubsection{Bias diagnostics and variance scaling} \label{sssec:sigmabias}

    \begin{figure}
        \centering
        \includegraphics[width=\columnwidth]{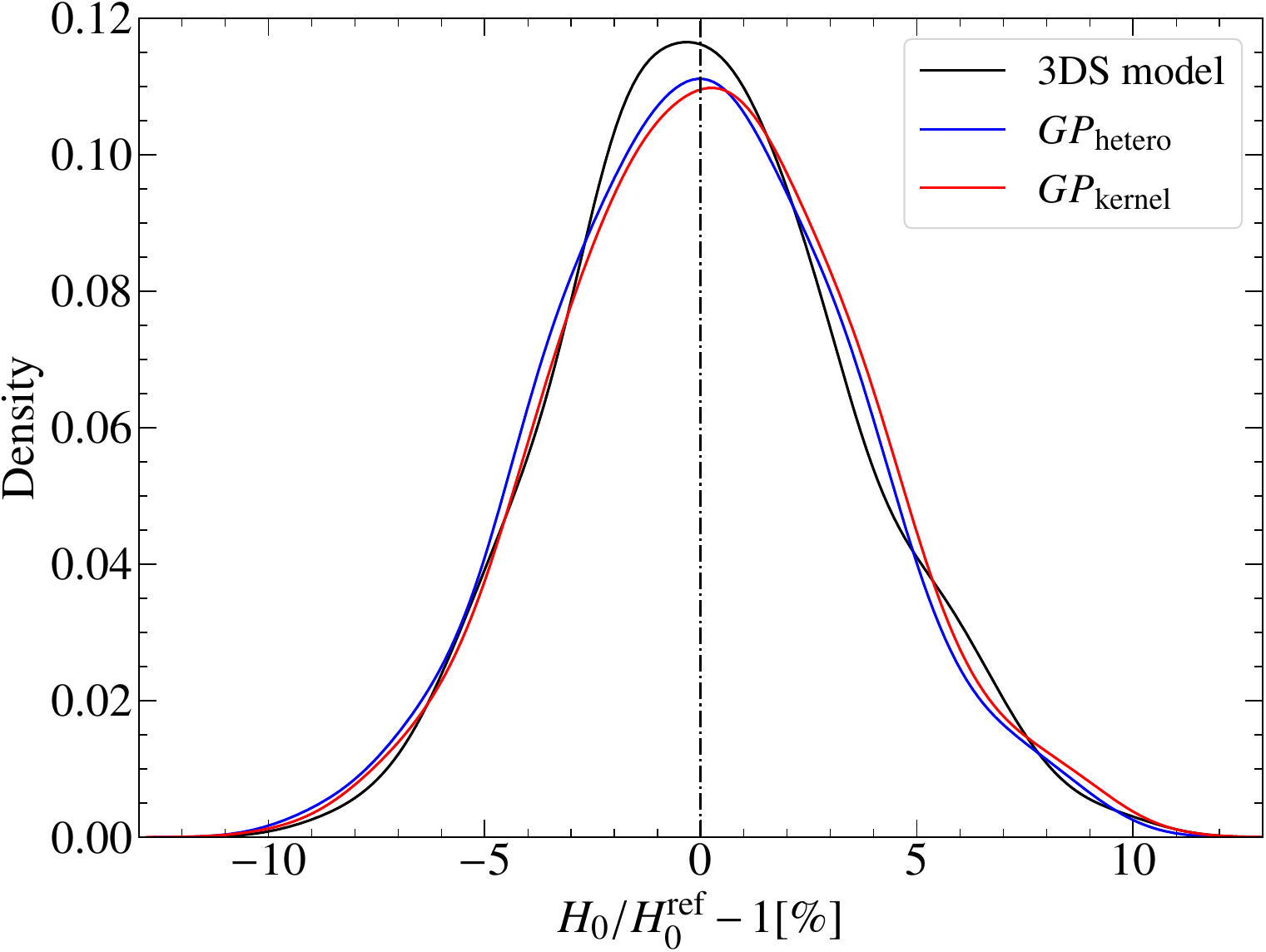} 
        \caption{Fractional differences in the $H_0$ MAP estimates with respect to the reference value adopted in the test for the 3DS (black), $GP_{\rm hetero}$ (blue), and $GP_{\rm kernel}$ (red) models of the informative priors, respectively.}
        \label{fig:1Atest}
    \end{figure}

\begin{figure}
        \centering
        \includegraphics[width=\columnwidth]{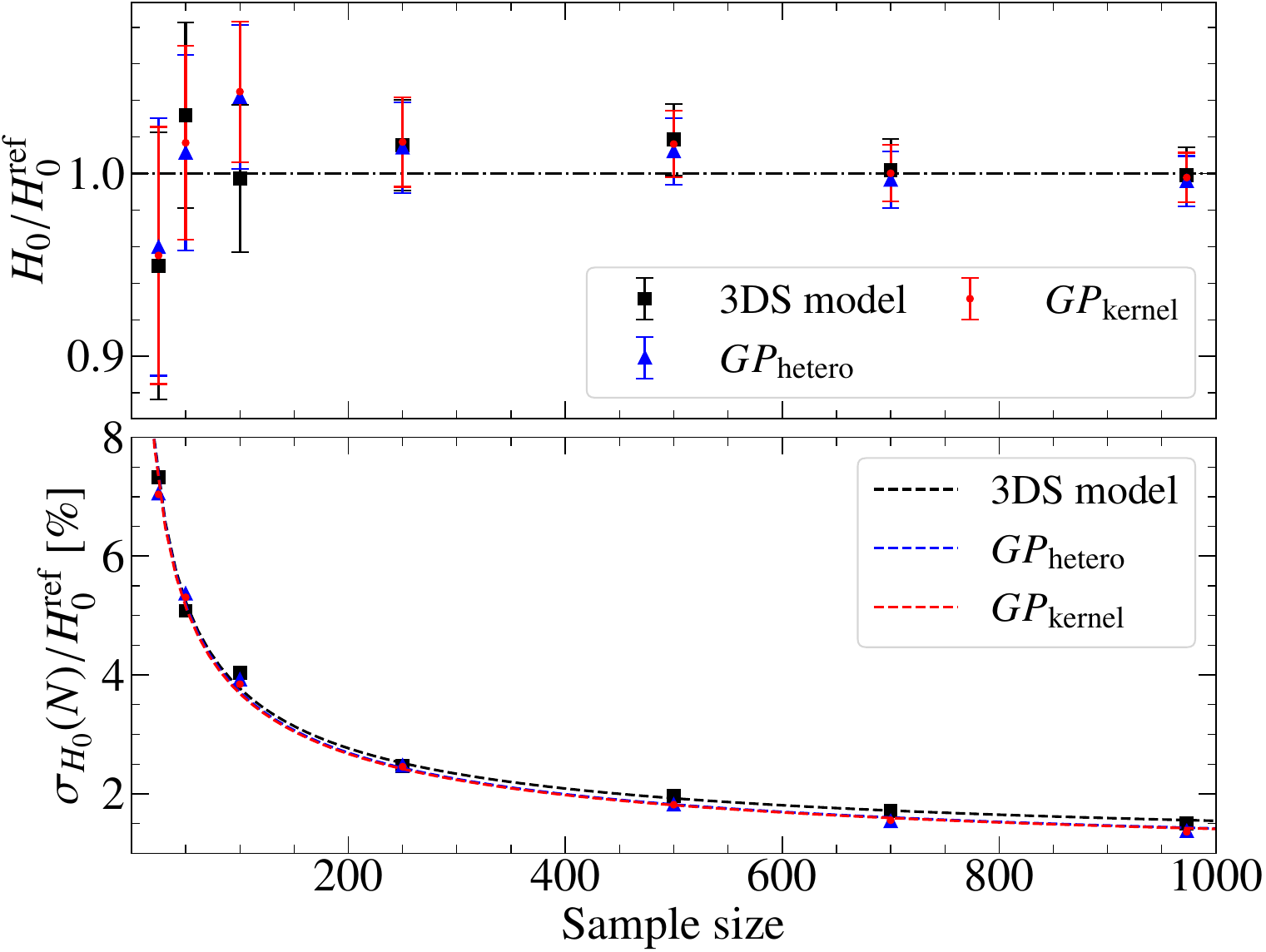} 
        \caption{Inferred $H_0$ values (upper panel) and its standard deviation (lower) as functions of the sample size. Black squares describe the results with the 3DS model, while blue triangles and red dots the $GP_{\rm hetero}$ and $GP_{\rm kernel}$ models, respectively. The three dashed lines show the fitted relations to the data for the three models.}
        \label{fig:Vartest}
    \end{figure}

    As a complementary test, we examined whether our inference produces biased maximum-a-posteriori (MAP) estimates, given the prior set-up of Table~\ref{tab:pr_Setup}. In particular, we generated $1000$ samples of $116$ clusters and measured the differences between the reference cosmological parameters and their MAP estimates. In Fig.~\ref{fig:1Atest}, we show the results for $H_0$. All three models for the informative priors yield similar results, with no significant biases in the MAP estimates (the mean bias is generally lower than $0.17\%$), while the dispersion is consistent with the variance expected for samples of that size, as shown in Fig.~\ref{fig:Vartest}.

    In particular, we implemented 10 chains of 5000 iterations, plus 1000 of tuning, for the MCMC with a Metropolis-Hastings sampler \citep[for a faster computation;][]{Metropolis1953, Hastings1970}, to study how the dispersion of the inferred $H_0$, and of their uncertainties, depends on sample size. By fitting a relation that includes a non-zero asymptotic (systematic) floor for the variance: $\sigma^2(N) = \sigma^2_0/N + \sigma^2_s$, we find similar values of $\sigma_0 \simeq 26.6$--$26.9 \,{\rm km\,s^{-1}Mpc^{-1}}$ for all models, while the systematic terms are small, but non-zero: $\sigma_s = 0.6$--$0.8 \,{\rm km\,s^{-1}Mpc^{-1}}$. The observed dispersion and the scaling of the standard deviations are therefore consistent with a regime dominated by statistical uncertainty for small samples, with relative uncertainties of the order of $4\%$ at about $100$ clusters and $1.5\%$ at approximately $1000$ clusters. For larger samples, however, the systematic term starts to dominate the relation. In particular, the statistical and systematic contributions become comparable at sample sizes of about $1225$ and $1850$ clusters for the 3DS and the two $GP$ models, respectively.

\subsubsection{Posterior calibration} \label{sssec:SBC}

    \begin{figure}
        \centering
        \includegraphics[width=\columnwidth]{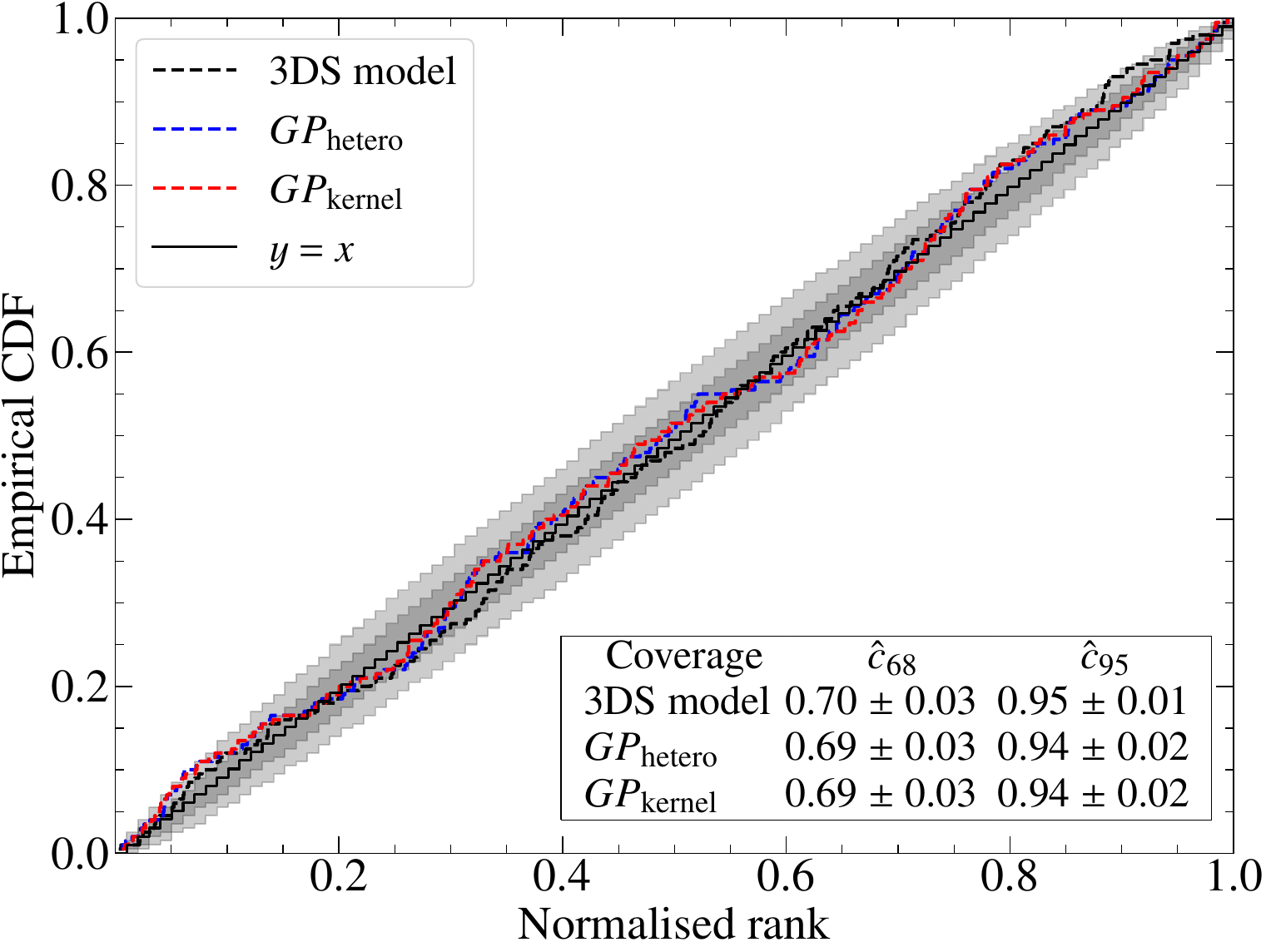} 
        \caption{Empirical cumulative distribution of the normalised ranks for the 3DS (black dashed line), $GP_{\rm kernel}$ (red), and $GP_{\rm hetero}$ (blue) models. The solid black line and the grey envelopes show the expectation for a uniform distribution and the confidence intervals at $68\%$ and $95\%$ around it. The coverage of the posteriors at $68\%$ and $95\%$ are also shown in the figure, for the three models.}
        \label{fig:SBC}
    \end{figure}

    As a final test, we studied whether the Bayesian analysis is well calibrated and posteriors have appropriate coverage. For this purpose, we implemented a simulation-based calibration \citep{Cook2006, Talts2018}. If the model and the inference procedure are correct, then by drawing parameters from the prior, simulating data from the likelihood, and inferring again the posteriors should recover the same joint distribution of parameters and data. More specifically, instead of adopting a fixed cosmology and a random selection of $\mathcal{B}$ values as detailed in Sect.~\ref{ssec:H0_valid}, we generated $200$ realisations of $\vec{\eta_T^{\rm sim}}$ by sampling parameters according to the priors listed in Table~\ref{tab:pr_Setup}: $\vec{\Theta}_{\rm sim} = (\vec{\vartheta}_{\rm sim}, \vec{\mathcal{B}}_{\rm sim}) \sim p(\vec{\Theta})$. With these parameters, we then simulated new data as $\vec{\eta_T^{\rm sim}} \sim p(\vec{\eta_T} \big\rvert \vec{\Theta}_{\rm sim})$ and drew new posteriors. In this way, $(\vec{\eta_T^{\rm sim}}, \vec{\Theta}_{\rm sim})$ is a draw from the joint distribution $p(\vec{\eta_T}, \vec{\Theta})$, which implies, for the Bayes theorem, that $\vec{\Theta}_{\rm sim}$ are drawn from the posteriors $p(\vec{\Theta} \big\rvert \vec{\eta_T^{\rm sim}})$ \citep{Cook2006}. If the inference is correctly implemented, the posterior samples and $\vec{\Theta}_{\rm sim}$ share the same marginal distribution, and the rank statistics of $\vec{\Theta}_{\rm sim}$ with respect to the posterior samples should be uniformly distributed. Deviations are instead symptoms of miscalibration, leading to biased posteriors or incorrect confidence intervals. In Fig.~\ref{fig:SBC}, we show the empirical cumulative distribution of the normalised ranks for $H_0$. For the three models, the cumulatives are consistent with uniformity, and the posteriors are adequately covered. In particular, within the $68\%$ and $95\%$ confidence intervals, the true cosmological parameter is recovered in $70 \pm 3\%$ and $95 \pm 1\%$ of the cases for the trichotomous model, and in $69 \pm 3\%$ and $94 \pm 2\%$ of the cases for the $GP$ models, respectively.

\section{Discussion} \label{sec:disc}

\subsection{Possible systematics and future perspectives} \label{ssec:sys}

    The main goal of this work was to extend and validate the methodology of \citetalias{Kozmanyan2019} in the controlled framework provided by \TheTH{} simulations. The results presented in Sect.~\ref{sec:res} show that simulation-informed priors on $\mathcal{B}$ can be consistently propagated into the Bayesian cosmological pipeline, yielding unbiased constraints on $H_0$ in mock tests. Future applications to observed clusters, however, may be affected by additional systematics related both to the accuracy of the hydrodynamical simulations and the quality of the observational data. A complete assessment of these effects is sample-dependent and beyond the scope of this simulation-only analysis. Nevertheless, below we briefly discuss some of these sources of uncertainty and how they could be addressed in future work.

    For the $GP$ models, the prediction does not consider uncertainties in the $\mathcal{M}_X$ parameter. However, this quantity can be affected by both statistical (e.g. Poisson noise) and observational systematic uncertainties (point source contamination or background and instrument modelling). To quantify the statistical contribution in the mock maps used in this work, we performed an additional Monte Carlo test. For each simulated X-ray count map, we generated Poisson realisations and recomputed the morphological indicators. With our mock maps, we found that the resulting $\mathcal{M}_X$ uncertainties are small, no larger than $\sigma_{\mathcal{M}_X} \lesssim 0.11$. This is well below the characteristic $GP$ correlation lengths, $\ell_\mu\simeq2$ for the mean trend and $\ell_\sigma\simeq3$ for the intrinsic scatter in the heteroscedastic $GP$ model. Since both the $GP$ mean and scatter vary smoothly on these scales, the statistical uncertainty on $\mathcal{M}_X$ is expected to have a subdominant impact in the present mock analysis. For real data, however, the uncertainties on $\mathcal{M}_X$ can be larger, depending on the quality of the X-ray observations, and could be propagated directly through the $GP$ prediction for $\mathcal{B}$ within the cosmological inference. Possible systematic biases in the morphological indicators would instead require a dedicated calibration based on more realistic mock observations of the observed sample under analysis.
    
    Hydrodynamical simulations necessarily provide an approximate description, whose level of accuracy depends on the physical processes and scales that they are designed to reproduce. In particular, the predicted ICM thermodynamic structure can depend on the adopted baryonic physics, especially in cluster cores. Within \TheTH{} the {\sc gizmo-simba} run, which was calibrated to better match the properties of the observed galaxies \citep{Cui2022}, presents higher ICM temperatures with relatively low gas densities at cluster centres compared to {\sc gadget-x} \citep{Li2023}. The present work is based on the {\sc gadget-x} implementation of \TheTH{}, which has been shown to successfully reproduce the observed gas properties (see, e.g. Sect.~\ref{sec:sim}) and their scaling relations \citep{Cui2018, Li2020}. Moreover, since $\mathcal{B}$ is inferred from relative mismatches between thermodynamic reconstructions, common-mode changes in the simulated profiles may partially cancel. Nevertheless, a residual dependence of the inferred $\mathcal{B}$ distribution on the adopted simulation model cannot be excluded.

    The $\mathcal{B}$ distribution studied in this work was derived from the $z=0$ snapshot and therefore does not explicitly test whether $\mathcal{B}$ evolves with redshift. Since scaled ICM thermodynamic profiles are generally found to evolve approximately self-similarly, at least at $z<1$ \citep[see, e.g.][and references therein]{Arnaud2010, McDonald2017, Ghirardini2019, Li2020}, a redshift evolution of $\mathcal{B}$ would be expected mainly if the fraction of dynamically disturbed systems changed significantly with redshift. In the first case, this effect can be captured, in first order, by re-weighting the class-conditional components of the 3DS model, or by adopting the appropriate morphology-dependent $GP$ priors. On the other hand, current observations do not provide clear evidence for an evolution of the cluster dynamical state at $z<1$, as the results depend on the adopted proxy and sample selection rather than reflecting an intrinsic evolution of relaxed and disturbed fractions \citep{Santos2010, Nurgaliev2017, Rossetti2017, McDonald2017, Bartalucci2019, Campitiello2022}. Moreover, \citet{DeLuca2026} found no clear redshift dependence for the observed $\eta_T$ values in the CHEX-MATE sample with $z < 0.6$. Our $\mathcal{B}$ distribution is also consistent with that of \citetalias{Kozmanyan2019}, which was derived from a simulated sample extracted at $z=0$, $0.25$, and $0.5$ and adopting a different cosmology (see the inset panel of Fig.~\ref{fig:Bdist}). Nevertheless, quantifying the residual dependence of $\mathcal{B}$ on the hydrodynamical solver, redshift, and possibly the assumed cosmological model requires more detailed studies based on multiple simulation runs and snapshots. In this respect, \TheTH{} provides a suitable framework for future work, since it includes the same simulated cluster regions with different hydrodynamical implementations.

\section{Summary and conclusions} \label{sec:conc}

    In this work, we used hydrodynamical simulations of galaxy clusters from \TheTH{} collaboration \citep{Cui2018} to characterise the cluster-structure bias, $\mathcal{B}$, and to evaluate its use as a simulation-informed prior in Bayesian cosmological analyses. For this purpose, we extended the methodology of \citetalias{Kozmanyan2019} to a larger synthetic sample, considering $324$ clusters from the {\sc gadget-x} simulation of \TheTH{} project, observed along three independent lines of sight. Our findings can be summarised as follows.
    
    The distribution of $\mathcal{B}$ is positively skewed and leptokurtic. A single log-normal distribution provides a good first-order description of the full sample, as expected for a quantity defined as a product of positive terms. However, $\mathcal{B}$ also depends on the dynamical and morphological state of the clusters. Relaxed systems are less scattered and lie closer to $\mathcal{B}\simeq 1$ than hybrid and disturbed clusters (Sections~\ref{ssec:Bdist} and~\ref{ssec:BvsDMS}). The full distribution is therefore better described by a dynamical class-conditional model, composed of three log-normal components, rather than a single log-normal population. This dependence can also be continuously modelled through morphology-dependent Gaussian-process regressions (Sect.~\ref{ssec:GP}).
    
    When these simulation-informed priors are incorporated into the cosmological analysis, the resulting Bayesian pipeline yields unbiased and well-calibrated posteriors in diagnostic mock tests assuming a $\Lambda$CDM model (Sect.~\ref{ssec:H0_valid}). The method is primarily sensitive to $H_0$, with a variance dominated by statistical uncertainties for small samples ($\lesssim 1000$ objects) and a non-zero asymptotic value of about $0.6$--$0.8 \,{\rm km\,s^{-1}Mpc^{-1}}$, which can be interpreted as the systematic limit for large ($\gtrsim 1200$) samples. The constraining power on the other cosmological parameters is limited, but combining this probe with cluster-based or other cosmological probes would be possible to help break parameter degeneracies. Therefore, our analysis shows that the structural term driving the discrepancy between X-ray and SZ thermodynamic reconstructions can be statistically characterised with hydrodynamical simulations and consistently incorporated into a Bayesian cosmological pipeline, supporting its future application to large X-ray and SZ cluster datasets. 
    
    Although these results were obtained with a specific hydrodynamical implementation, the {\sc gadget-x} run of \TheTH{}, the inferred distribution of $\mathcal{B}$ is consistent with previous simulation-based results and observational trends \citep[\citetalias{Kozmanyan2019};][]{DeLuca2026}. Moreover, as discussed in Sect.~\ref{ssec:sys}, this simulation has been shown to reproduce the observed gas properties and thermodynamic profiles \citep{Cui2018, Li2020, Rossetti2024}. Future extensions of this work can address additional observational and simulation-based sources of uncertainty, further improving the robustness of the simulation-informed priors and their use in cosmological analyses.

\begin{acknowledgements}
    
    We thank the anonymous referee and the editors for their useful comments. The authors acknowledge The Red Española de Supercomputación for granting computing time for running the hydrodynamical simulations of \TheTH{} galaxy cluster project in the Marenostrum supercomputer at the Barcelona Supercomputing Center. HB, FDL, and PM acknowledge the financial contribution from the contract Prin-MUR 2022 supported by Next Generation EU (M4.C2.1.1, n.20227RNLY3 \textit{The concordance cosmological model: stress-tests with galaxy clusters}) and the support by INFN through the InDark initiative. GY, WC, and DdA were partially supported by grant PID2024-156100NB-C21, funded by MICIU/AEI/10.13039/501100011033 and by “ERDF A way of making Europe” (European Union). WC further acknowledges Comunidad de Madrid for the Atracci\'{o}n de Talento fellowship no. 2020-T1/TIC19882 and Agencia Estatal de Investigaci\'{o}n (AEI) for the Consolidaci\'{o}n Investigadora Grant CNS2024-154838, EU and ERC: HORIZON-TMA-MSCA-SE for supporting the LACEGAL-III (Latin American Chinese European Galaxy Formation Network) project with grant number 101086388 and the science research grants from the China Manned Space Project. ER is supported by NASA EGIP Grant 80NSSC25K0006 and 80NSSC25K8009. MDP acknowledges support from Sapienza Università di Roma thanks to Progetti di Ricerca Medi 2023, prot. RM123188F4731D3E. FDL, HB, and PM also thank the INFN Roma2 IT group for their invaluable support, particularly in the aftermath of the recent fire at their facilities. \\
    This work used diverse \texttt{python} packages: \texttt{numpy} \citep{numpy}, \texttt{scipy} \citep{SciPy}, \texttt{matplotlib} \citep{matplot}, \texttt{seaborn} \citep{seaborn}, \texttt{pandas} \citep{pandas, pandas2}, \texttt{astropy} \citep{astropy,astropy2,astropy3}, \texttt{scikit-learn} \citep{scikit-learn}, \texttt{pymc} \citep{pymc2023}, \texttt{arviz} \citep{arviz2019}, \texttt{getdist} \citep{getdist}.
    
\end{acknowledgements}

\bibliographystyle{aa}
\bibliography{biblio}
\clearpage

\begin{appendix}

\section{Dependence on LOS integration} \label{sec:los}

    \begin{figure}[h]
        \centering
        \includegraphics[width=\columnwidth]{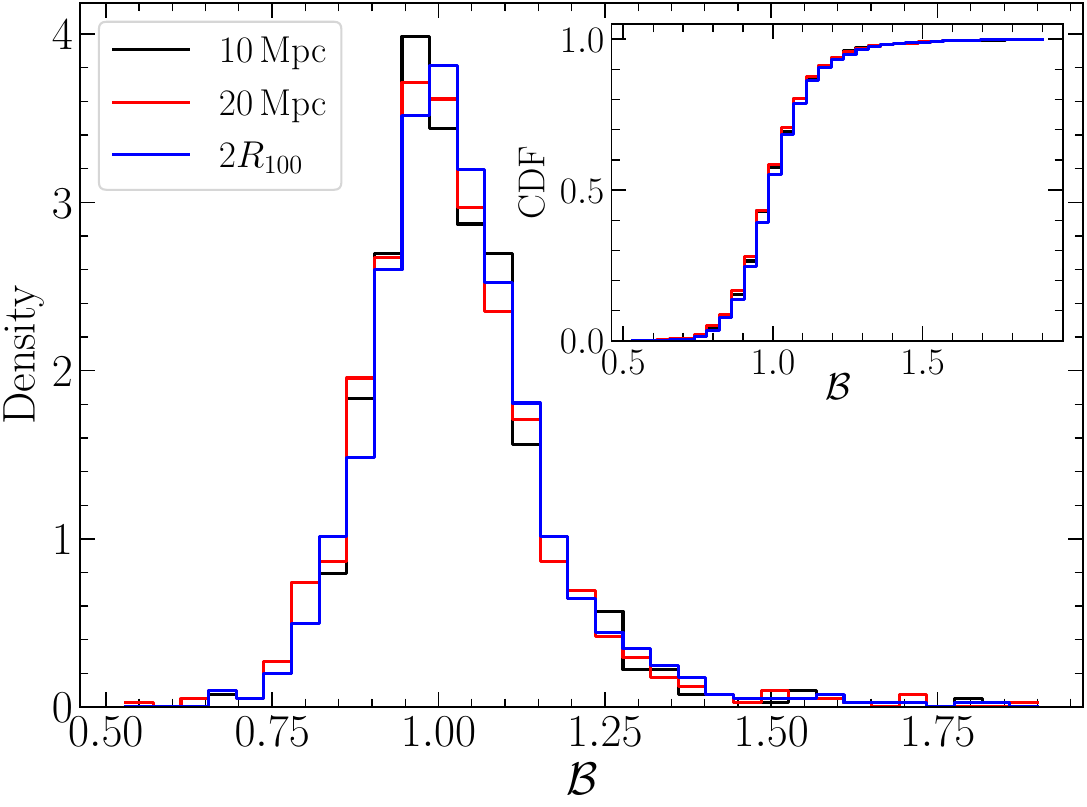} 
        \caption{Histograms and cumulative (inset panel) distributions of the $\mathcal{B}$ parameter calculated with integration lengths of $10\,\rm Mpc$ (black), $20\,\rm Mpc$ (red), and $2R_{100}$ (blue).}
        \label{fig:eta_integ}
    \end{figure}

    \noindent
    In the evaluation of the discretised quantities described by Eqs.~\eqref{eq:EM},~\eqref{eq:Tsl}, and~\eqref{eq:y_sim}, an important factor is the integration length (i.e. the height of the cylinder along the LOS). Generally, if the size of the cluster is known, the integration length can be related to a characteristic cluster scale. Otherwise, a common choice is to set the lengths large enough to collect the entire cluster signal. To test whether the choice of a specific integration length impacted the results (e.g. due to gas particles along the LOS not gravitationally bounded to the cluster, such as those within filaments), we compare in Fig.~\ref{fig:eta_integ} the $\mathcal{B}$ distribution obtained with typical integration lengths of $2R_{100}$, $10 \,\rm Mpc$, and $20 \,\rm Mpc$ \citep[e.g.][]{Meneghetti2010, Rasia2012, Rasia2015, Kozmanyan2019}. The three distributions are remarkably similar, with only minor differences in the case of the smallest integration length used, $2R_{100}$. More specifically, comparing the distributions with the KS test, we cannot reject the null hypothesis (that the two samples are drawn from the same parent distribution) since the $p$-values are significantly larger than the typical significance thresholds of $5\%$ or $1\%$. For the test between $10 \,\rm Mpc$ and $20 \,\rm Mpc$, we have a $p$-value of $0.89$, while for $2R_{100}$, the $p$-values are $0.36$ with $10 \,\rm Mpc$ and $0.58$ for $20 \,\rm Mpc$.

\section{Bias in $\mathcal{B}$ from repeated clusters in the sample} \label{sec:Bvsproj}

\begin{figure}
        \centering
        \includegraphics[width=\columnwidth]{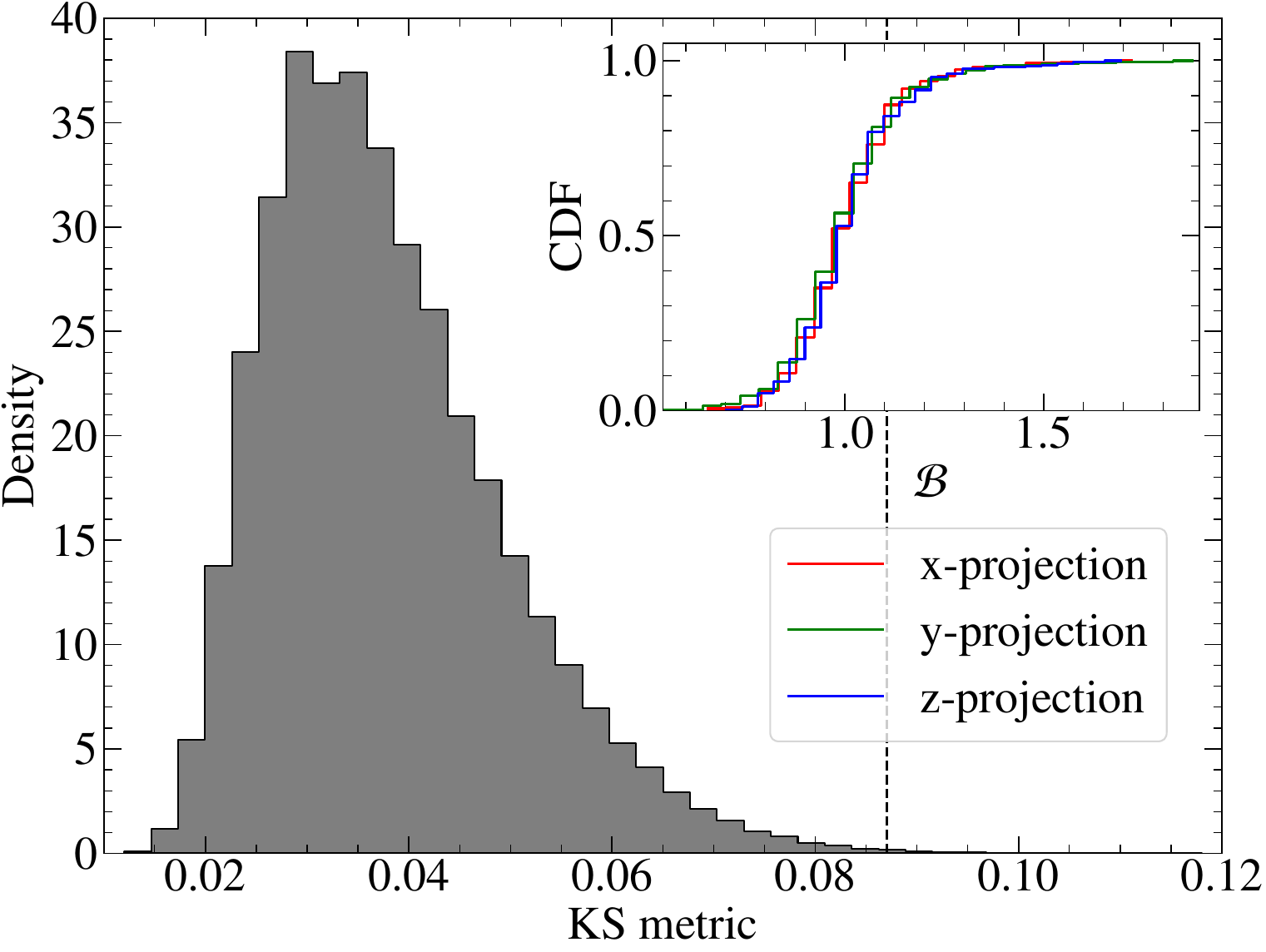}
        \caption{Distribution of the KS statistic from the Monte Carlo realisations. The vertical line shows the threshold in the KS metric corresponding to a $p$-value of $0.05$. The inset panel shows the cumulative distributions of the $\mathcal{B}$ for the three different projections.}
        \label{fig:proj_test}
    \end{figure}

    In this work, we built our sample by collecting $\mathcal{B}$ values from three different projections of $324$ clusters, considering the principal axes of the simulation cubes. When the three projections are analysed separately, their $\mathcal{B}$ distributions do not present significant differences, as shown from their cumulatives in the inset panel of Fig.~\ref{fig:proj_test}. Moreover, two-sample KS tests between each pair of projections yield $p$-values larger than $0.50$. Nevertheless, this similarity does not guarantee that different projections of the same haloes can be used as statistically independent clusters. To test whether the inclusion of repeated clusters introduces biases in our analysis, we performed a Monte Carlo resampling, in which one projection per cluster was randomly selected to build subsamples of $324$ systems, similarly to \citetalias{Kozmanyan2019}. In particular, we generated $10^6$ subsamples and compared the resulting $\mathcal{B}$ distributions with the full one using KS tests. If a significant fraction of the realisations yielded KS statistics for which the null hypothesis could be rejected, then the effect of including repeated clusters could not be neglected. The distribution of the KS metric for the Monte Carlo realisations is shown in Fig.~\ref{fig:proj_test}, together with the threshold corresponding to a $p$-value of $0.05$ as a vertical dashed line. Only for less than $0.1\%$ of the realisation we can reject the null hypothesis. Moreover, we do not observe systematic deviations in the mean, median, or variance of the Monte Carlo realisations with respect to the overall distribution. Therefore, the inclusion of repeated halos does not introduce significant biases in the analysis.

\section{Morphological indicators} \label{sec:Morp_300}

\begin{figure*}[p]
        \sidecaption
        \includegraphics[width=12cm]{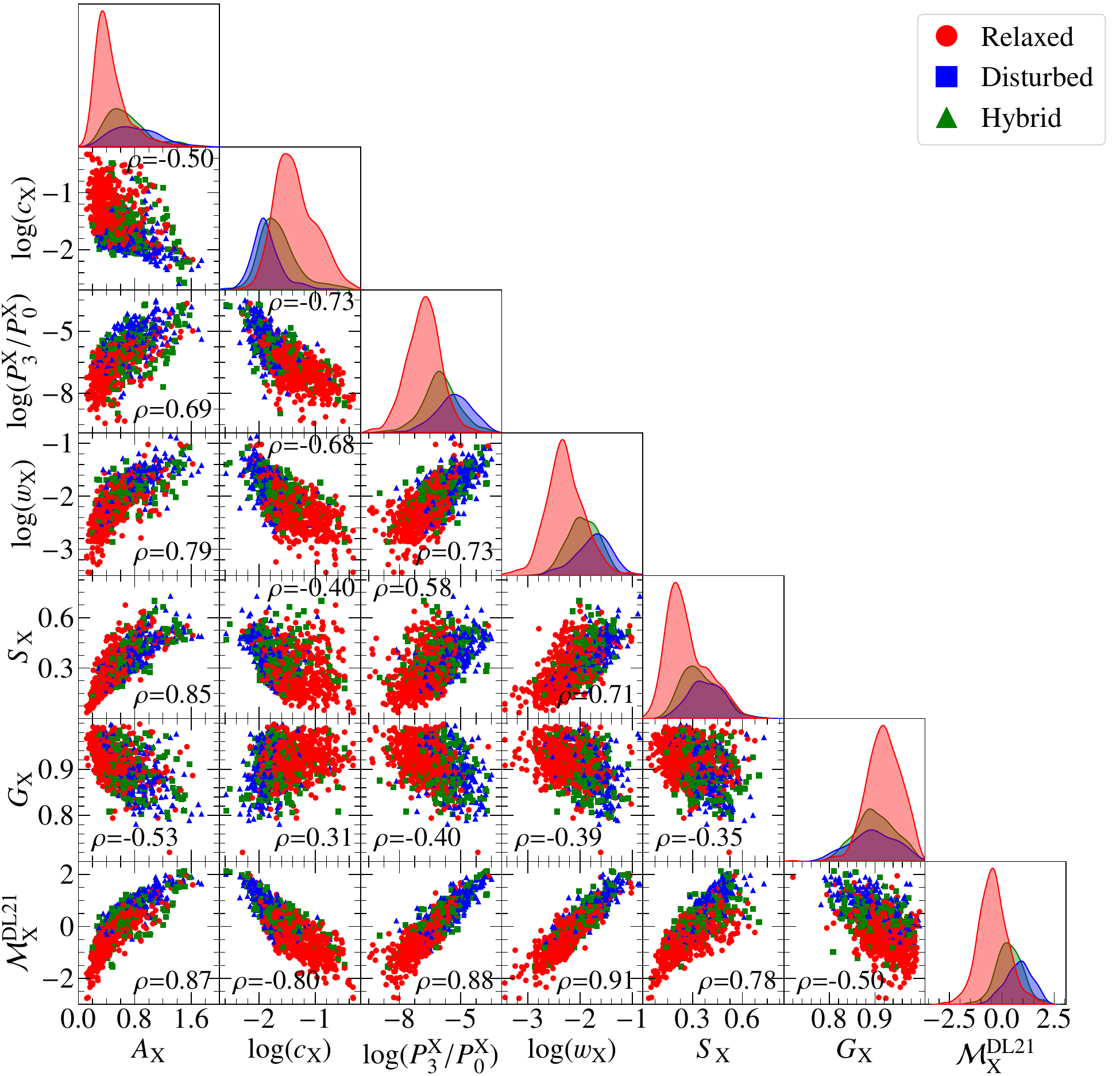}       
        \caption{Correlation between the X-ray morphological indicators as defined in \citetalias{DeLuca2021}. The Spearman correlation coefficients, $\rho$, are reported in each panel. Relaxed, hybrid, and disturbed clusters are shown with red, green, and blue, and are marked by circles, triangles, and squares, respectively.}
        \label{fig:2DX_corr}
    \end{figure*}

\begin{figure*}[p]
        \sidecaption
        \includegraphics[width=12cm]{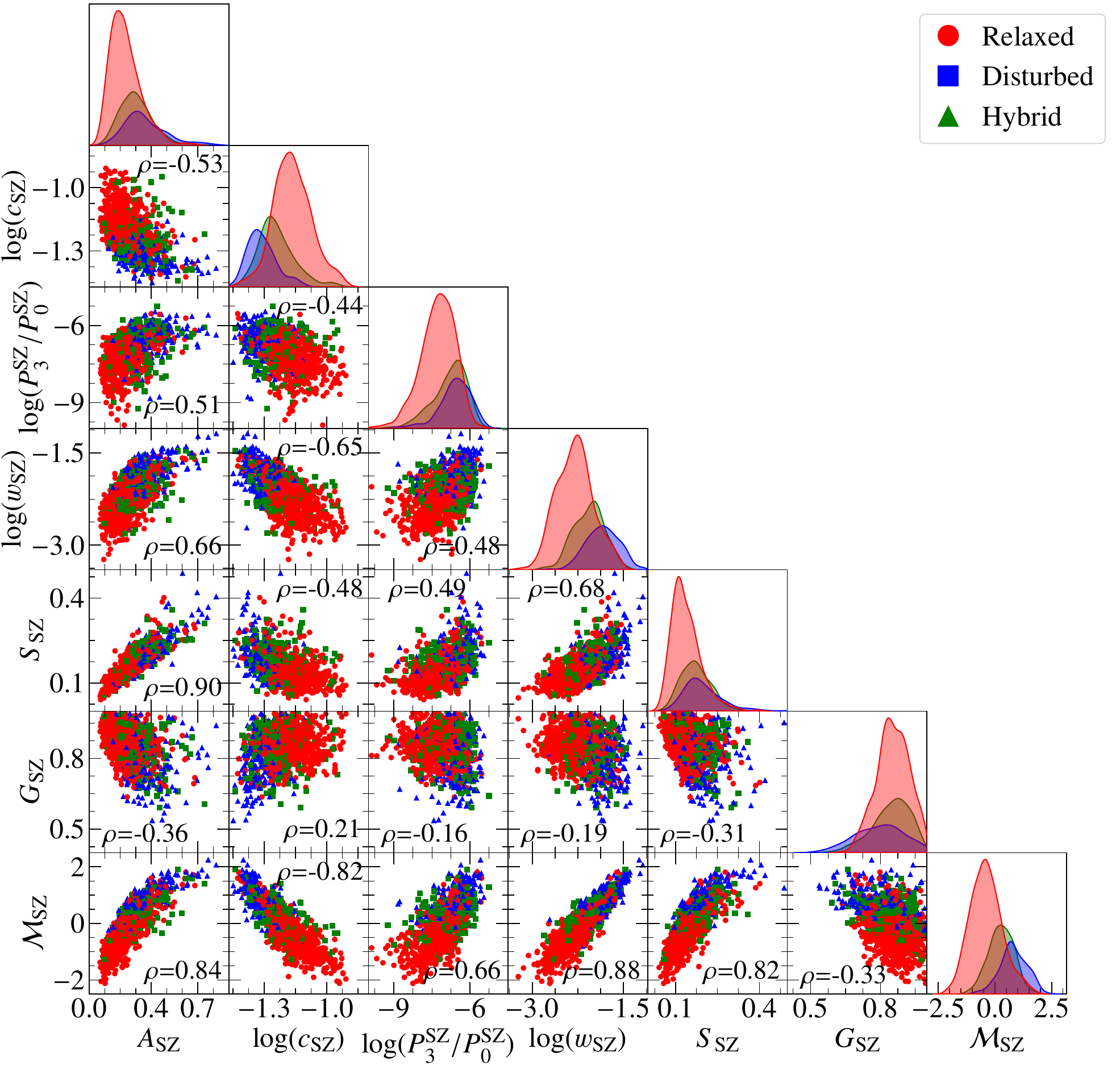} 
        \caption{Correlations between the SZ morphological indicators, considering the definition of \citetalias{DeLuca2021}. The same colour code of Fig.~\ref{fig:2DX_corr} is used for relaxed (red), hybrid (green), and disturbed (blue) systems. The Spearman correlation coefficient, $\rho$, is also reported in each panel.}
        \label{fig:2DSZ_corr}
    \end{figure*}

    In Figs.~\ref{fig:2DX_corr} and~\ref{fig:2DSZ_corr}, we show the distributions of the morphological indicators for the X-ray and SZ maps following the definition of \citetalias{DeLuca2021}. In particular, we used the same maps, centres, apertures, and weights. As already discussed by \citetalias{DeLuca2021}, the results obtained from the X-ray and SZ maps are generally similar, although the correlations between the indicators differ. Parameters designed to capture large-scale inhomogeneities (e.g. $A$ or $S$) are more effective in SZ maps, since the signal is approximately proportional to $n_\mathrm{e}$. Instead, the X-ray emission is approximately proportional to $n_\mathrm{e}^2$, and thus indicators more sensitive to small-scale variations, such as $c$ and $w$, show stronger correlations in X-ray maps. The combined indicators are strongly correlated with almost all parameters in both X-ray and SZ, as a result of the tuning procedure described in \citetalias{DeLuca2021}. The weakest parameter in tracing cluster morphologies is $G$, which shows only mild correlations with the other indicators. Therefore, $\mathcal{M}$ provides a good summary indicator for ranking cluster morphologies, as it captures the main features traced by each individual indicator.

    \afterpage{\clearpage}
    
\begin{figure*}
        \sidecaption
        \includegraphics[width=12cm]{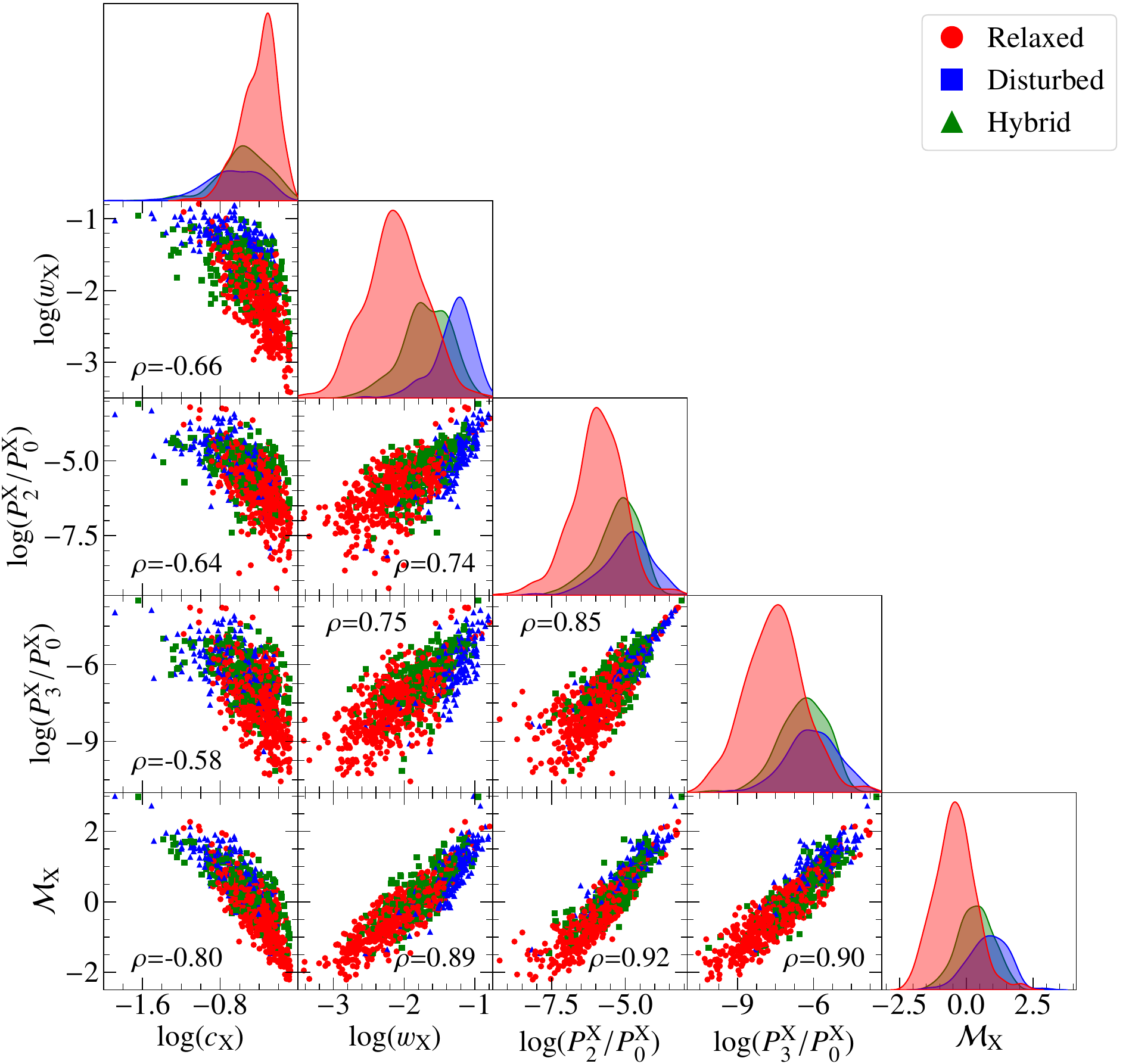} 
        \caption{Distribution of the X-ray morphological indicators described in Sect.~\ref{ssec:morph}. Relaxed, hybrid, and disturbed clusters are shown in red, green, and blue, respectively. The Spearman correlation coefficients, $\rho$, are reported in each panel.}
        \label{fig:2DX4par_corr}
    \end{figure*}

    For comparison, we show in Fig.~\ref{fig:2DX4par_corr} the X-ray morphological indicators used in this work, based on the parameter definitions adopted by \citet{Campitiello2022}. In fact, the study of \citetalias{DeLuca2021} was based only on simulated datasets and aimed at identifying the optimal apertures and weights for the indicators. However, in order to build informative priors that can be readily applied to real observations, our models should follow as closely as possible the methodology already adopted for observed clusters. For this reason, we considered the results presented in \citet{Campitiello2022} for the CHEX-MATE clusters. In particular, we smoothed the X-ray maps to mimic observations at the angular resolution of \XMM{}, and then applied the morphological indicators as described in Sect.~\ref{ssec:morph}. As a general remark, the indicators selected by \citet{Campitiello2022} are among the most effective X-ray morphological parameters, and their mutual correlations remain broadly similar to those shown in Fig.~\ref{fig:2DX_corr}.

    \FloatBarrier

\section{Prior significance} \label{sec:Priors}

    The cosmological pipeline described in Sections~\ref{ssec:BtoH0} and~\ref{ssec:H0_valid} is based on a single observable, $\eta_T$, and hence different constraining powers and parameter degeneracies are expected among the inferred cosmological parameters. In this Appendix, we discuss the results of the test about the sensitivity of our Bayesian inference to the choice of the cosmological priors. 

    \begin{figure}
        \centering
        \includegraphics[width=\columnwidth]{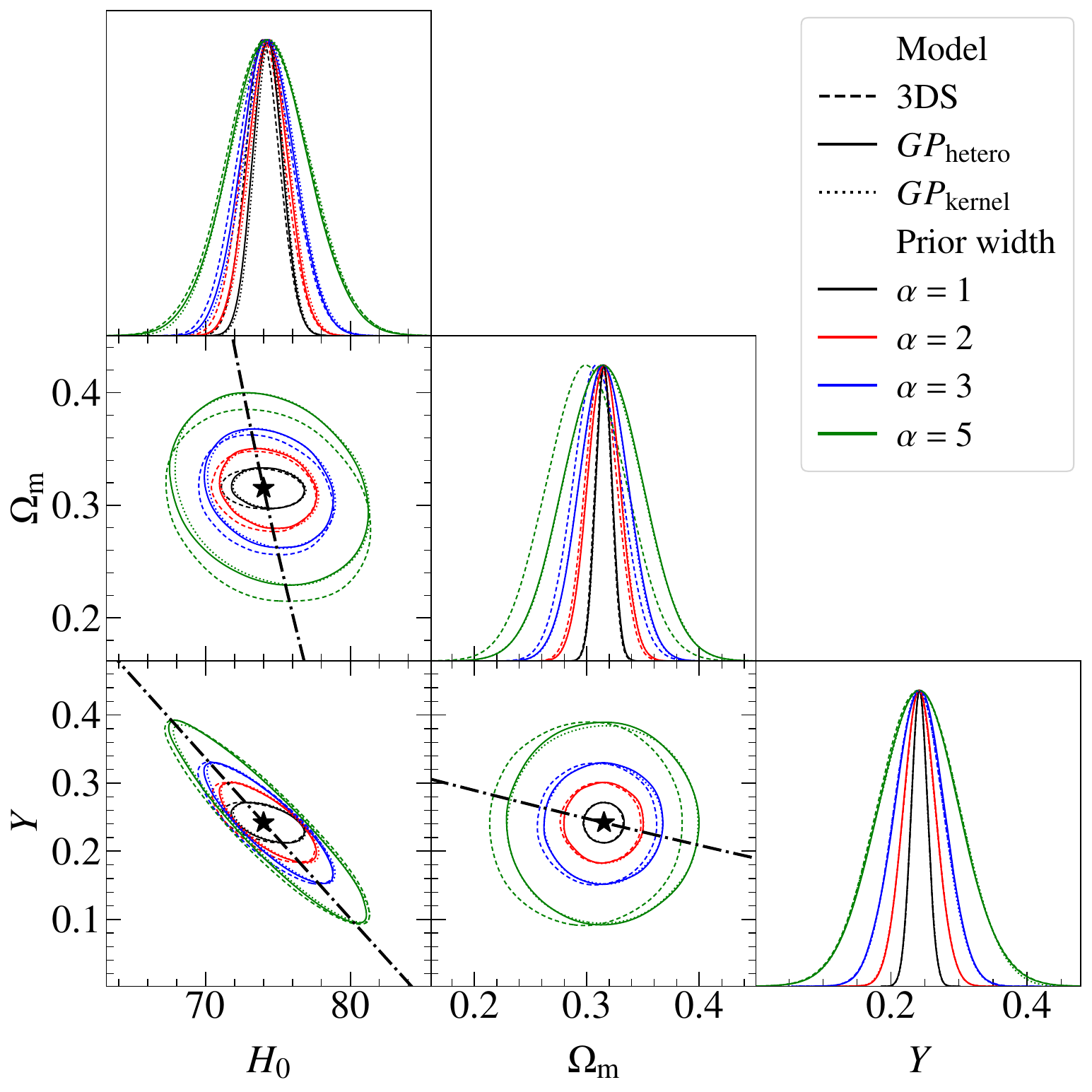} 
        \caption{Posterior distributions on the cosmological parameters obtained by relaxing the prior on $\Omega_\mathrm{m}$ and $Y$ by a factor $\alpha$. Dashed lines are used to describe the result of the 3DS model, while solid and dotted lines for $GP_{\rm hetero}$ and $GP_{\rm kernel}$ models, respectively. The star markers show the reference cosmological values used in the test, and the dash-dotted curves the theoretical degeneracy expected between the parameters. The contours enclose $95\%$ of the joint posterior probability.}
        \label{fig:Pr_corner}
    \end{figure}
    
    As a first test, we progressively relaxed the Gaussian priors on $\Omega_\mathrm{m}$ and $Y$ by increasing their standard deviations by a factor $\alpha = (1, 2, 3, 5)$. The resulting marginal posteriors are shown in Fig.~\ref{fig:Pr_corner}. While the Hubble constant remains well constrained by the data, the posteriors of $\Omega_\mathrm{m}$ and $Y$ are largely driven by the priors. More quantitatively, the rescaled width ratio $(\sigma_{\rm prior}/\sigma_{\rm posterior})/\alpha$ for $Y$ is consistent with unity for all values of $\alpha$, while, for $\Omega_\mathrm{m}$, we find values of $0.95$ and $0.98$ for $\alpha=3$ and $5$, respectively, and values close to unity in the other cases.
    
    As discussed in Sect.~\ref{sssec:priors}, $\eta_T$ is less sensitive to the angular diameter distance ($\eta_T \propto D_\mathrm{A}(z)^{-1/2}$) than other probes, such as $f_{\rm gas}$, which scales as $f_{\rm gas} \propto D_\mathrm{A}(z)^{3/2}$ \citep[see e.g.][]{Allen2011}. Using only $\eta_T$ and with realistic uncertainties, the degeneracies between the parameters cannot be fully broken, as shown by the joint posteriors of Fig.~\ref{fig:Pr_corner}. The most evident one is between $H_0$ and $Y$, as also noted by \citet{Ettori2020}, since both act qualitatively as normalisation factors in Eq.~\eqref{eq:cosm_biases}. The expected relation between the parameters can be derived from Eqs.~\eqref{eq:cosm_biases},~\eqref{eq:Dfunc}, and~\eqref{eq:Yfunc}. At fixed $\Omega_\mathrm{m}$ and keeping constant $\mathcal{C}$, we can write:
    \begin{equation}
        \frac{H_0(Y)}{H_{0}^{\rm ref}} = \left[\frac{\mathcal{Y}(Y_{\rm ref})}{\mathcal{Y}(Y)}\right]^2 = \left( \frac{2+4Y_{\rm ref}\xi}{2-Y_{\rm ref}+2\xi} \right) \left( \frac{2+4Y\xi}{2-Y+2\xi}\right)^{-1}.
        \label{eq:H0Y}
    \end{equation}
    Considering that $\xi$ is generally small (of the order of $\sim 10^{-3}$) and $Y<1$, the expected degeneracy is almost linear in this regime, with an approximate negative slope of $-1/(2-Y_{\rm ref})$.

    The degeneracy between $H_0$ and $\Omega_\mathrm{m}$, at fixed $Y$, can be derived instead from the definition of the angular diameter distance. In particular, the expected scaling is given by:
    \begin{equation}
        \frac{H_0(\Omega_\mathrm{m})}{H_{0}^{\rm ref}} = \frac{A(z;\Omega_\mathrm{m})}
         {A(z;\Omega_\mathrm{m}^{\rm ref})},
         \label{eq:H0Om}
         \quad
         A(z;\Omega_\mathrm{m}) = \int_0^z \frac{{\rm d}z'}{E(z',\Omega_\mathrm{m})} .
    \end{equation}
    For small variation around a reference value $\Omega_\mathrm{m}^{\rm ref}$, Eq.~\eqref{eq:H0Om} can be linearised as:
    \begin{equation}
        \frac{H_0(\Omega_\mathrm{m})}{H_{0}^{\rm ref}} \simeq 1 - \frac{\Omega_\mathrm{m}-\Omega_\mathrm{m}^{\rm ref}}{2A(z; \Omega_\mathrm{m}^{\rm ref})} \int_0^{z} \frac{\partial_{\Omega_\mathrm{m}} E^2(z',\Omega_\mathrm{m}^{\rm ref})}{E^3(z',\Omega_\mathrm{m}^{\rm ref})} \mathrm{d}z', 
        \label{eq:H0Om_approx}
    \end{equation}
    with, for example, $\partial_{\Omega_\mathrm{m}}E^2=(1+z)^3-1$ considering a flat $\Lambda$CDM model with a negligible radiation term. Equations~\eqref{eq:H0Om} and~\eqref{eq:H0Om_approx} were derived for a single cluster. For cluster samples, the scaling can be evaluated by averaging $H_0(\Omega_\mathrm{m})/H_{0}^{\rm ref}$ over all clusters, or adopting an effective redshift. In any case, the partial derivative $\partial A/\partial \Omega_\mathrm{m}$ is negative, and thus higher $\Omega_\mathrm{m}$ values correspond to lower $H_0$ values. In Fig.~\ref{fig:Pr_corner}, the expected trends are shown in the joint posterior distributions with dash-dotted curves. In particular, when the prior is relaxed (e.g. for the case $\alpha=5$), the joint posterior between $H_0$ and $Y$ in Fig.~\ref{fig:Pr_corner} clearly follows the expected behaviour given by Eq.~\eqref{eq:H0Y}.
    
\begin{figure}
        \centering
        \includegraphics[width=\columnwidth]{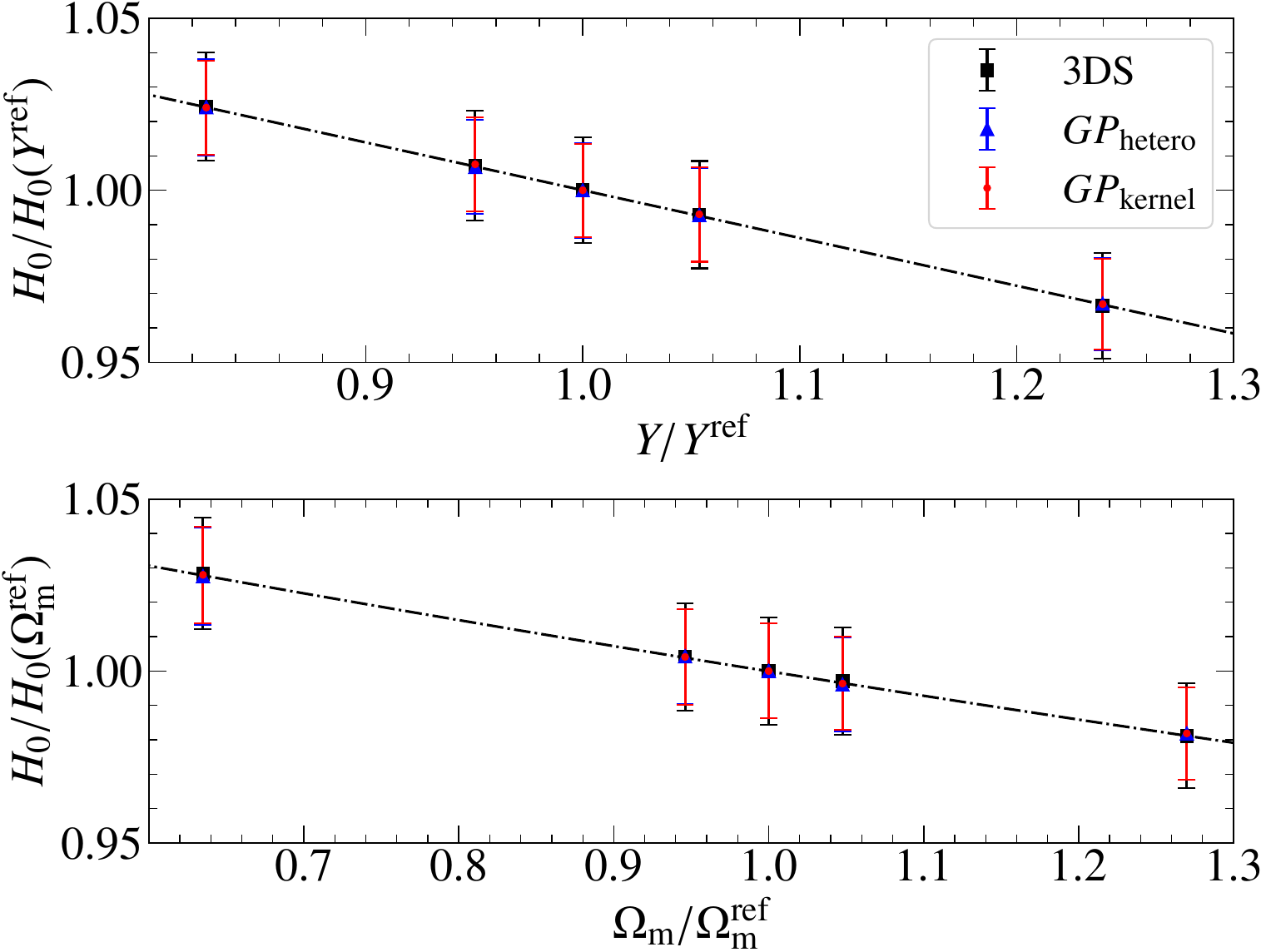} 
        \caption{Variations in the inferred $H_0$ values obtained by shifting the prior central values of $Y$ (upper panel) and $\Omega_\mathrm{m}$ (lower) with respect to the reference values. Black squares show the results for the 3DS model, while blue triangles and red circles the $GP_{\rm hetero}$ and $GP_{\rm kernel}$ models, respectively. The black dash-dotted curves show the expected trends from the analytical parameter degeneracies of Eqs.~\eqref{eq:H0Y} and~\eqref{eq:H0Om}.}
        \label{fig:Pr_shift}
    \end{figure}

    As a second complementary test, we quantified how shifts in the assumed central values of $\Omega_\mathrm{m}$ and $Y$ priors can affect the inferred value of $H_0$. In particular, we repeated the inference on the same mock catalogues using the reference prior widths (i.e. with $\alpha=1$), but varying one prior centre at a time. The results are shown in Fig.~\ref{fig:Pr_shift}, together with the scaling described by Eqs.~\eqref{eq:H0Y} and~\eqref{eq:H0Om}. For the tested values, the recovered $H_0$ follows the expected trends for all three models. Shifts of about $5\%$ in either $Y$ or $\Omega_\mathrm{m}$ induce changes smaller than $1\%$ in $H_0$, while even for the most extreme variations considered here, of about $20$--$35\%$, $H_0$ changes at most of $\sim 3.4\%$. Therefore, any dependence on the assumed prior centres is mainly driven by the cosmological factor $\mathcal{C}$, rather than by the adopted model for $\mathcal{B}$. 
    
    To conclude, we note that the shifts explored here are deliberately conservative. Current determinations of primordial helium abundance generally reach the percent-level, although mild tensions among recent measurements have been reported \citep[e.g. see the review of][]{DiValentino2025}. Similarly, $\Omega_\mathrm{m}$ is constrained at the percent level by current CMB and baryon acoustic oscillation analyses, although its central value may depend on the adopted dataset combination and cosmological model \citep[e.g.][]{DESI-DR2-2025}. In this context, the induced shifts in $H_0$ remain sub-dominant with respect to the statistical uncertainties considered here. A more detailed assessment of the impact of alternative cosmological priors is therefore deferred to future applications of the method to observed cluster samples.

\end{appendix}

\end{document}